\documentclass[bookmarks=true, twocolumn]{aastex6}

\usepackage{amssymb,amsmath,eurosym}
\usepackage{mathrsfs}
\usepackage{enumerate}
\usepackage{natbib}
\newcommand{\figref}[1]{Figure \ref{#1}}
\newcommand{\vect}[1]{\mathbf{#1}}

\newcommand{\fnc}[1]{\textsf{#1}} 
\newcommand{\unit}[1]{\ensuremath{\, \mathrm{#1}}}
\newcommand{\del}{\nabla}

\providecommand{\abs}[1]{\lvert#1\rvert}
\providecommand\natexlab[1]{#1}
\providecommand\JournalTitle[1]{#1}

\newcommand{\change}[1]{{#1}}

\DeclareMathOperator{\sech}{sech}

\defcitealias{Tarr:2017a}{Paper I}

\begin{document}
\title{The formation and dissipation of current sheets and shocks due to compressive waves in a stratified atmosphere containing a magnetic null}
\author{Lucas A. Tarr\altaffilmark{1,2}}
\affil{George Mason University, Dept. of Physics and Astronomy, 4400 University Drive, Fairfax, VA 22030}
\author{Mark Linton}
\affil{Naval Research Laboratory \\
4555 Overlook Ave SW \\
Washington, DC 20375, USA
}

\altaffiltext{1}{ltarr2@gmu.edu}
\altaffiltext{2}{A portion of this work was performed while LAT held an NRC postdoctoral position at NRL.}

\begin{abstract}
  We study the propagation and dissipation of magnetohydrodynamic waves in a set of numerical models that each include a solar--like stratified atmosphere and a magnetic field with a null point.  All simulations have the same magnetic field configuration but different transition region heights.  Compressive wave packets introduced in the photospheric portion of the simulations refract towards the null and collapse it into a current sheet, which then undergoes reconnection.  \change{The collapsed null forms a current sheet due to a strong magnetic pressure gradient caused by the inability of magnetic perturbations to cross the null.  Although the null current sheet undergoes multiple reconnection episodes due to repeated reflections off the lower boundary, we find no evidence of oscillatory reconnection arising from the dynamics of the null itself}.  Wave mode conversion around the null generates a series of slow mode shocks localized near each separatrix.  \change{The shock strength is asymmetric across each separatrix, and subsequent shock damping therefore creates a tangential discontinuity across each separatrix, with long--lived current densities.  A parameter study of the injected wave energy to reach the null confirms our previous WKB estimates.}  Finally, using current estimates of the photospheric acoustic power, we estimate that the shock and Ohmic heating we describe may account for $\approx1-10\%$ of the radiative losses from \change{coronal} bright points with similar topologies, and are similarly insufficient to account for losses from larger structures such as ephemeral regions.  At the same time, the dynamics are comparable to proposed mechanisms for generating type--II spicules.
\end{abstract}

\section{Introduction}
In our \change{previous} paper \citep[][hereafter \citetalias{Tarr:2017a}]{Tarr:2017a} we reported on a numerical simulation of magnetohydrodynamic (MHD) wave propagation through a model solar atmosphere.  The study of MHD waves in stratified atmospheres stretches back many decades (see for example \citet{Ferraro:1958} and the extensive bibliography given in \citetalias{Tarr:2017a}).  Many observations have demonstrated the ubiquity of oscillatory phenomena in basically every feature of the solar atmosphere that researchers have decided to study: sunspot umbra and penumbra, pores, active region filaments, large prominences, quiet sun regions, extreme ultraviolet (EUV) loops above active regions, and the more extended corona \change{\citep[see the Introduction in \citetalias{Tarr:2017a} for specific examples, or reviews by][]{Nakariakov:2005,Khomenko:2013,Khomenko:2015, Warmuth:2015, Arregui:2018}.}  These observations indicate that a substantial amount of wave energy is present throughout the atmosphere and therefore this energy is likely important for solving the long--standing coronal and chromospheric heating problems.  This general topic raises the important question: ``How do the waves get to the locations where they are observed?''  Convective motions in the photosphere are the obvious source, but early models found that convectively induced compressive waves should reflect from relatively low in the atmosphere, or even below the photosphere for lower frequency waves.  An important step forward was the recognition that mode conversion---the transfer of wave energy between different types of MHD waves---allows MHD waves to propagate to locations that are inaccessible to their hydrodynamic cousins.  This topic was the focus of \citetalias{Tarr:2017a}.

A distinguishing feature of our previous work was that the magnetic field contained a strong inhomogeneity, owing to the presence of a single magnetic null point.  At the null, the magnetic field is zero and has a discontinuous first derivative.  \change{Although MHD waves in the presence of nulls have been studied repeatedly in the past \citep[see the Introduction and Discussion in \citetalias{Tarr:2017a}, and reviews by][]{McLaughlin:2011, Pontin:2012}, few had done so in the context of a stratified atmosphere.}  The presence of the null turned out to greatly affect the propagation of waves---and the energy they carry---throughout the domain, compared to the case of a more smoothly varying field.  The null created a location in the upper chromosphere where mode conversion could take place, in addition to the lower--lying conversion region typically considered\change{, for example, in studies of mode conversion in sunspots \citep{Rijs:2016}}.  Essentially, the null allows wave energy to tunnel through a region in which it would otherwise be evanescent, although this process happens in a roundabout way involving mode conversion.

\change{Wave mode conversion (the exchange of energy between the fast, Alfv\'en, and slow MHD waves, though we focus here only on the fast and slow waves), is an important process in our simulations, and we have discussed it at length in \citetalias{Tarr:2017a}.  Here, we summarize several important concepts needed to understand the present work.  Mode converstion may occur when the eigenvectors describing the MHD modes become nearly degenerate \citep{Tracy:2003}.  A necessary condition for this is that the sound and Alfv\'en speed are equal, $c_s=v_A$.  At such locations, there is near equipartition between pressure and magnetic forces, which, for a $\gamma=5/3$ plasma, give plasma $\beta=2c_s^2/\gamma v_A^2 = 1.2$.  Hence, regions where $\beta\approx 1$ allow for mode conversion.  Magnetic null points have $\vect{B}=0$, meaning that (i) $v_A\rightarrow0$ at nulls, (ii) a high--$\beta$ region must exist around a null, and therefore, (iii) any wave approaching a null is susceptible to mode conversion (except in the strictly $\beta=0$ limit).  Further, this means the nature of an MHD wave may change considerably as it crosses an equipartition region and transitions from a high--$\beta$ to low--$\beta$ environment, or vice versa.  At the same time, the properties of fast and slow waves are only truly distinct far from equipartition layers, which complicates the discussion of mode conversion and wave propagation near the equipartition layer, and around nulls in particular.  See \citet{McLaughlin:2006b} for an early study of, and attempt to quantify, mode conversion near a magnetic null.  We refer the reader to \citetalias{Tarr:2017a}, especially their \S5.1 and \S7.2, for further discussion on mode conversion near nulls and how the properties of fast and slow MHD waves change across an equipartition boundary.}

In \citetalias{Tarr:2017a} we found that the conversion process at the null coincided with two important dynamic effects: (i) the null point collapsed into a current sheet, and (ii) a set of propagating slow waves arose at the conversion region surrounding the null.  The formation, oscillation, and dissipation of the current sheet suggested that the incoming wave induced magnetic reconnection at the null point.  Indeed, this appears to be a general feature of the interaction of waves with magnetic nulls: waves refract toward and deposit energy at null points, thereby forming current sheets.  In the other causal direction, whenever reconnection happens, the unbalancing of the Lorentz force generates MHD waves, and the newly launched waves may steepen to form shocks as they propagate away from the reconnection site, heating the plasma away from the null in the process.  In this sense, the study of MHD waves and reconnection are not separate topics, and will often need to be treated simultaneously.  In our previous work we only noted the existence and oscillation of the current sheets and slow mode waves.  In the present work we will discuss how the slow modes form shocks, how the current sheet forms, and the ultimate fate of both.

In \citetalias{Tarr:2017a} we did not investigate the physical process by which the current sheet arose and then oscillated.  However, \citet{Murray:2009} explored what appears to be a similar phenomenon in their 2.5D simulation of an emerging flux tube in a background field.  A null point naturally formed between the emerging and surrounding flux, and as emergence drove the system, the null periodically collapsed into a current sheet and underwent reconnection.  It is unclear if the current sheet forms in the same way in our wave driving experiment, and we answer the question of how our current sheet forms, below.  The salient feature of \citet{Murray:2009} is that the current sheet retracts because force balance across separatrices near the outflow causes the separatrices to pull apart.

\change{We describe the present simulations in \S\ref{sec:sim}, including, in \S\ref{sec:setup}, how they differ from our previous work in \citetalias{Tarr:2017a}, and the primary results in \S\ref{sec:primary_results}.  Details for one representative simulation are presented in \S\ref{sec:formation}, which explains the null point collapse in terms of force balance in \S\ref{sec:CSformation}, quantifies the reconnection rate and compares it to Ohmic heating in \S\ref{sec:rx}, and explains slow shock formation and dissipation in \S\ref{sec:shocks}.  Part one of the parameter study, in \S\ref{sec:longtermprops}, compares the long term properties for simulations with varying transitions region heights, $y_{tr}$.  Part two of the parameter study, in \S\ref{sec:param_study}, determines the mode conversion efficiency as a wavepacket is introduced at different locations throughout the simulation, and confirms the results of a WKB analysis given in \citetalias{Tarr:2017a}.  Section \S\ref{sec:discussion} applies our results to the question of radiative losses in coronal bright points, and compares our results with other relevant work.  Finally, we briefly summarize our results in \S\ref{sec:conclusion}.}

\section{Simulation Overview}\label{sec:sim}
\subsection{Setup and differences from \citetalias{Tarr:2017a}}\label{sec:setup}
\begin{figure}
  \includegraphics[width=0.45\textwidth]{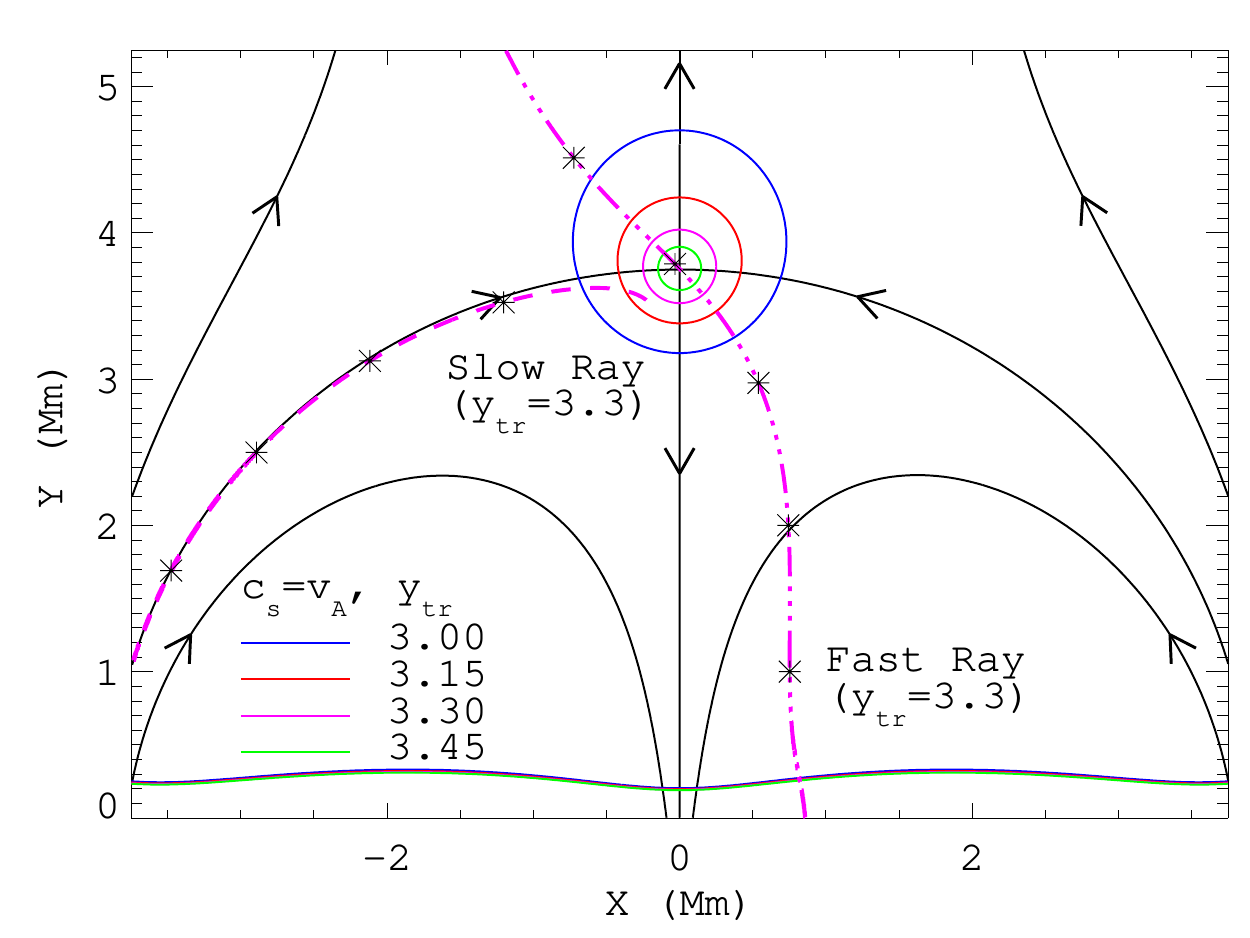}
  \caption{\label{fig:multi-csva} Representative magnetic field lines (black with arrows) from the initial condition in the high resolution region.  Blue, red, magenta, and green solid lines show equipartition contours where $c_s=v_A$ as the transition region height is increased (and thus the density at the null drops).  Blue corresponds to the simulation studied in \citetalias{Tarr:2017a}.  Magenta corresponds to the simulation with $y_{tr}=3.3\unit{Mm}$, which is the primary focus of this paper.  The dash and dash--dotted lines are the trajectories of a slow and fast ray traversing the $y_{tr}=3.3\unit{Mm}$ simulation, respectively, and $*$ symbols mark $1\unit{Mm}$ intervals along each ray.}
\end{figure}

The simulations studied in this work are updated versions of the one described in \citetalias{Tarr:2017a}; we refer the reader to \S2 of that paper for a complete description, and here note only changes made to better isolate the behavior of the shocks.  The full resistive MHD equations are solved using the LARE2D code \citep[version 2.11;][]{Arber:2001} \change{in terms of the primitive variables mass density, specific internal energy, plasma velocity, and magnetic field, $\rho, \epsilon, \vect{v}, \vect{B}$, respectively}.  Our simulation consists of an initially magnetohydrostatic atmosphere that is subsequently driven by injecting a compressive wave packet through the lower boundary.  The solid black lines with arrows in \figref{fig:multi-csva} show the magnetic configuration near the null, and the solid contours show several cases of the plasma stratification, as described below.  The null is located at $(x,y)=(0,3.75)\unit{Mm}$, and the four field lines connected to it form the separatrix dome.  Compared to \citetalias{Tarr:2017a}, we introduce three main changes to the simulation: two to the initial condition and one change to the wave driver.  

First, we now use a stretched numerical grid that has high resolution in regions of interest near the null and lower resolution approaching the side and top boundaries.  The high resolution region has uniform grid spacings $\Delta_x=\Delta_y\equiv\Delta=L_N/32\approx4.7\unit{km}$ and spans $\abs{x}\lesssim9\unit{Mm}$ horizontally and $0<y\lesssim6\unit{Mm}$ vertically.  We use the subscript ``N'' to denote numerical normalization factors, which are unchanged from \citetalias{Tarr:2017a}.  The factor $L_N=150\unit{km}$ normalizes the numerical lengths, and is the density scale height at the model photosphere.  The high resolution is a factor of four (in each direction) greater than that used in our previous simulation.  The other numerical normalizations are $\rho_N=3.03\times10^{-4}\unit{kg}\unit{m}^{-1}$, which normalizes the density, and $B_N=0.12\unit{T}$ which, normalizes the magnetic field.  All remaining normalizations can be derived in terms of $B_N, L_N, \rho_N$.  For quick reference, velocity is normalized to $v_N=B_n/\sqrt{\mu_0\rho_N}\approx6.177\unit{km/s}$, time to $t_N=L_N/v_N \approx 24.28\unit{s}$, and current to $j_N=B_N/(\mu_0L_N)\approx 0.63\unit{A}\unit{m}^{-2}$.  \change{Resistivity is normalized to $\mu_0 L_N v_N$, so that $S=\mu_0L_Nv_N/\eta$ is the Lundquist number defined using the pressure scale height at the lower boundary.  The uniform resistivity is $\eta=0.0333\approx \Delta/L_N$, compared to a value of $0.1$ in \citetalias{Tarr:2017a}\footnote{\change{We followed the same guidelines in setting the uniform resistivity $\eta$ as we did in \citetalias{Tarr:2017a}.  See their \S2.1 for a discussion of how numerical resistivity scales with the grid resolution and typical lengthscales of dynamic evolution.}}.  For the remained of the paper, we have written equations in normalized units, yet restored physical units for the figures}.

We stretch the grid in both the vertical and horizontal directions using $\tanh$ profiles as described in Appendix \ref{sec:stretch}.  The uniform, high resolution region is described above.  The lowest resolution regions (furthest from the dome) have $\Delta_x \approx 0.526L_N\approx 79\unit{km}$ and $\Delta_y\approx0.47L_n\approx70\unit{km}$.  The stretched grid ensures that the side and top boundaries are far from the null point and injection locations, mitigating possible boundary effects, such as reflections.  The side and top regions also include the damping term described in \citetalias{Tarr:2017a}, which reduces reflections by removing kinetic energy near the boundaries.  The full simulation has $6144\times4096$ computation cells and spans $\abs{x}\lesssim52.5\unit{Mm}$ and $0<y<120\unit{Mm}$ in the horizontal and vertical directions, respectively.

For the second main difference, we modify the height of the transition region $y_{tr}$, varying from the value $y_{tr}=3.0\unit{Mm}$ used in \citetalias{Tarr:2017a} (see their Equation 7) to $y_{tr}=3.45\unit{Mm}$ in steps of $0.15\unit{Mm}$.  As the transition region height increases the stratification results in a lower (and more realistic) coronal density: the plasma temperature, and through it the density scale height, remains lower to a greater height above the photosphere, and therefore the density decreases.  \change{The primary simulation we focus on has $y_{tr}=3.3\unit{Mm}$; compared to \citetalias{Tarr:2017a}, the density decreases an additional $\approx 2$ $e$--foldings before stabilizing to an isothermal corona with nearly uniform density.}

\change{In all simulations} we have kept the null point at a height of $y_{null}=3.75\unit{Mm}$ so that the density is equal to (for $y_{tr}=3.0\unit{Mm}$) or lower at the null compared to the simulation in \citetalias{Tarr:2017a}.  For increasing $y_{tr}$, the sound speed is correspondingly lower near the null while the Alfv\'en speed is higher, so that the the equipartition layer, where $c_s=v_A$, is situated closer to the null, with an approximate radius of $r_E\approx 0.25\unit{Mm}$ for the $y_{tr}=3.3\unit{Mm}$ simulation.  The solid green, magenta, red, and blue contours in \figref{fig:multi-csva} show (for, respectively, the $y_{tr} = 3.45, 3.30, 3.15, 3.0\unit{Mm}$ simulations) the two equipartition regions in the simulation, one around the null and one running near the lower boundary (where all the curves nearly overlap).  When used without qualification, ``equipartition region'' refers to the one surrounding the null.  \figref{fig:multi-csva} shows how the equipartition region shrinks as the transition region height increases and density at the null decreases.  The magenta curve corresponds to the primary simulation in this work, with $y_{tr} =3.3\unit{Mm}$.  The blue contour corresponds to the \citetalias{Tarr:2017a} case, where $r_{E}\approx0.75\unit{Mm}$.  We will briefly compare the results of simulations using all four values $y_{tr}$ in \S\ref{sec:longtermprops}.

For the third main difference, we modify the introduced ``wave'' to contain just a single Gaussian--shaped pulse (c.f. \citetalias{Tarr:2017a} Equation 21).  This was done to better study a single interaction of the pulse and the null point.  When a wave interacts with a null it generates a complicated response that includes current sheet formation, mode conversion, and shock generation.  If subsequent amplitude peaks in a wave train are introduced they further interact with each type of response, complicating the analysis.  For this reason, in this paper, we have chosen to focus on the system's response to just a single pulse.

We introduce a wave pulse using the method described in \citetalias{Tarr:2017a}: we specify a spatio--temporal vertical velocity perturbation in the simulation's lower--boundary ghost cells and use the velocity perturbation to define adiabatic energy and density perturbations.  The velocity pulse is described in space and time by
\begin{equation}
  \label{eq:vdrive}v_y(x,y,t) = v_d\exp\Bigl[-\frac{(x-x_d)^2}{2w_x^2}-\frac{(y-c_st)^2}{2w_y^2}\Bigr]
\end{equation}
with associated specific internal energy and density terms
\begin{equation}
  \label{eq:de-drive}\rho_1 = \rho_0v_y/c_s  \qquad\text{and}\qquad \epsilon_1 = \rho_1(\gamma-1)\frac{\epsilon_0}{\rho_0}.
\end{equation}
Subscripts $0$ and $1$ refer to background and perturbation quantities, respectively.  \change{For the primary simulation,} the pulse has spatial parameters $x_d=1\unit{Mm}$, $w_x=1.977L_N=0.297\unit{Mm}$, and $w_y=0.15L_N=0.0225\unit{Mm}$, and is advected upward through the lower boundary at the local sound speed $c_s=\sqrt{\gamma P_0/\rho_0}=\sqrt{\gamma(\gamma-1)\epsilon_0(y=0)}$.  The terms $x_d$ and $w_x$ are the pulse's centroid and width in $x$.  Because of the vertical advection at the sound speed, the vertical width $w_y$ sets the driving duration $T_d\approx w_y/c_s$.  \change{In the parameter study discussed in \S\ref{sec:param_study}, we vary the wavepacket injection location $x_d$ between $-6.0\unit{Mm}$ and $3.0\unit{Mm}$ for the $y_\text{tr}=\unit{Mm}$ atmosphere.}

\begin{figure*}
  \begin{center}
    \includegraphics[width=0.95\textwidth]{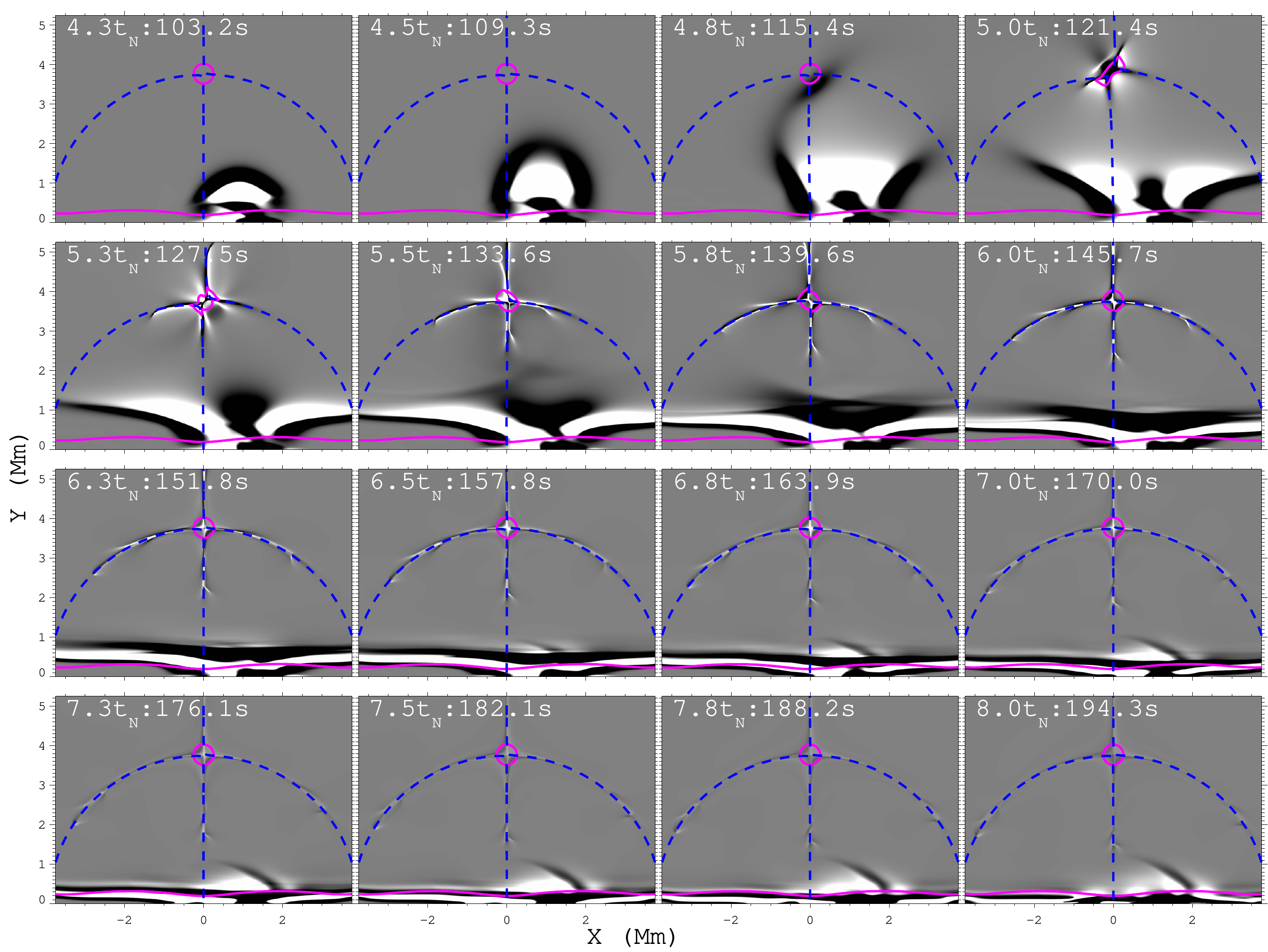}
    \caption{\label{fig:jzfull} Evolution of the out--of--plane current density $j_z$ during the $y_{tr}=3.3\unit{Mm}$ simulation.  Magenta shows the equipartition contour, and the separatrix field lines are shown in dashed blue.  The grayscale is saturated at $\pm 10^{-3}J_N \approx 6.4\times 10^{-4}\unit{A}\unit{m}^{-2}$.}
  \end{center}
\end{figure*}

To better isolate the waves from the background, we also run a baseline simulation, with the exact same initial condition, but no injected wave.  The difference between the simulations with and without the wave isolates the wave--perturbed quantities $\rho_1$, $\vect{v}_1$, etc.  Other methods of background subtraction, such as differencing the wave--injected simulation with itself at an initial time or a running average of previous times, all show nearly the same dynamics, but also include minor contributions from currents and velocities that arise due \change{to} secular drifts in the simulation caused by imperfect numerical representation of derivatives \citep[see][\S.2.2]{Tarr:2017a}.  These dynamics are unimportant compared to those of the injected wave packet: amplitudes in the baseline simulation are typically at least 2 orders of magnitude less than those in the wave injection simulation.  The background subtraction we use primarily produces visualizations with fewer distracting features when the color scales are saturated to best visualize the dynamics near the null, but does not otherwise affect the analysis.

\subsection{Primary results}\label{sec:primary_results}
\figref{fig:jzfull} shows a time series of the $y_{tr}=3.3\unit{Mm}$ simulation output in terms of the out--of--plane current density.  This view highlights the essential elements of the numerical experiment: (i) the pulse, injected at $\vect{x}_0=(1,0)\unit{Mm}$, initially rises and expands ($t<4.5t_N$); (ii) a portion of the pulse refracts towards the null while the majority refracts away and back to the photosphere ($t\approx4.8 t_N$); (iii) the null point collapses into a current sheet ($t\approx 5.3t_N$); (iv) the current sheet eventually retracts and reforms oriented $90^\circ$ to the original sheet, and with opposite current density ($t\approx 5.8t_N$); (v) outward propagating disturbances from the null are largely confined to be near the separatrices ($t\ge5.3t_N$); and finally, (vi) some additional oscillations at the null take place as the system seeks a (new) equilibrium state.

\begin{figure*}[ht]
  \begin{center}
    \includegraphics[width=0.99\textwidth]{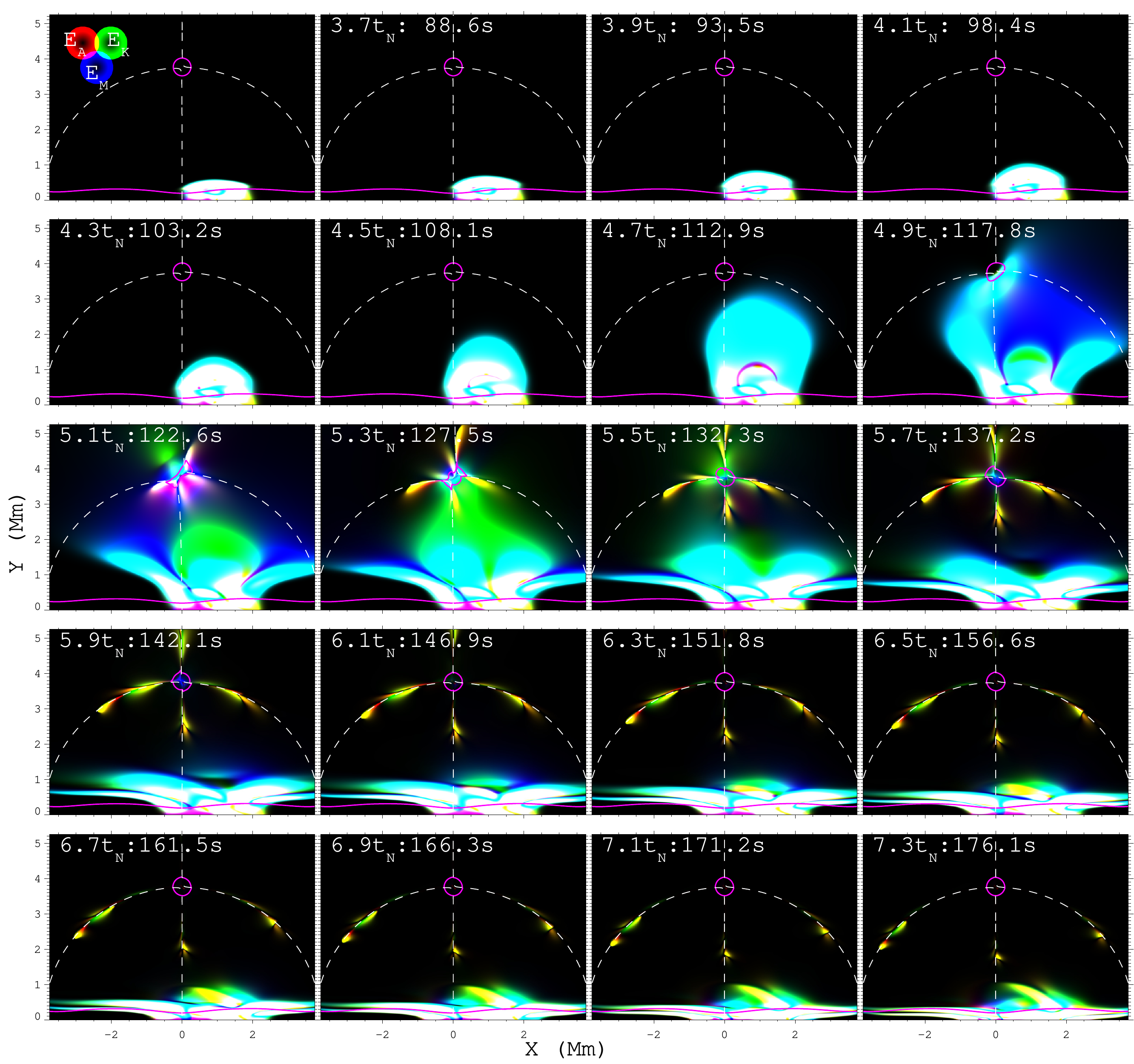}
    \caption{\label{fig:enfull} Time sequence of energy densities for acoustic (red), kinetic (green), and magnetic (blue) terms from Equation \eqref{eq:wave-conservation}.  Dashed lines trace out the separatrix field lines.  The two $c_s=v_A$ curves are shown in magenta.  The color wheels in the upper left show the overlap of different energy densities, and are on a linear scale from $0$ at each center to $1.1\times 10^{-2}\unit{J}\unit{m}^{-2}$ at the radii where the circles intersect, and then constant beyond that.  An animation of this figure is available in the online material (\texttt{ycor22\_p1\_waveE.mp4}).}
  \end{center}
\end{figure*}

Other physical parameters (velocity, density perturbation) show essentially the same information as the current density in \figref{fig:jzfull}.  A more useful representation of the data for understanding the dynamics is shown in \figref{fig:enfull}.  Here, each color channel represents a different energy density term in the wave conservation relation derived in \citetalias{Tarr:2017a},
\begin{multline}
  \label{eq:wave-conservation}\partial_t\Biggl[\frac{1}{2}\rho_0v_1^2+\frac{\change{P}_1^2}{2\rho_0c_s^2}+\frac{\abs{\vect{b}_1^2}}{2}\Biggr] \\
  +\del\cdot\Bigl[\change{P}_1\vect{v}_1+(\vect{B}_0\times\vect{v}_1)\times\vect{b}_1\Bigr]=0.
\end{multline}
The three terms within the time derivative are the kinetic ($E_K$), acoustic $(E_A)$, and magnetic ($E_M$) energy densities, and the two terms within the divergence are acoustic $(\vect{F}_A)$ and magnetic ($\vect{F}_M$, i.e. Poynting) fluxes.  Note that each term is a second order quantity that describes an energy density or energy flux of waves, excluding the background.  The upper left panel contains three color wheels that show how the terms combine.  Cyan (yellow) colors show equipartition between kinetic and magnetic (acoustic) energy densities.  Therefore, outside the magenta contours in \figref{fig:enfull}, where $\beta<1.2$, cyan (yellow) represent fast (slow) waves, while inside the magenta contours ($\beta>1.2$) the reverse holds.  The color scales are saturated to best represent the wave energy densities in the upper chromosphere to low corona, basically around the height of the null\footnote{\change{The relations between the acoustic versus magnetic partition of wave potential energy, the type of MHD wave, and the local value of $\beta$ is described in detail in \citetalias{Tarr:2017a}: see their \S3 and \S{5.1}.}}.

\figref{fig:enfull} and its animation (available in the online material) illustrate the dynamics of the simulation in terms of the \change{partition of wave energy ($E_K,\ E_A,\ E_M$) at each location}.  Initially the pulse travels upward ($t=3.5-4.7t_N$) and then splits ($\approx4.9t_N$) into multiple pieces.  The sides refract back toward the photosphere while the central region refracts toward the null.  All wave sections are predominately fast mode waves in the $\beta\lesssim1$ regions\change{, having near equipartition between magnetic and kinectic energies (cyan)}.  The splitting of the initial wave packet into multiple subpackets was described in detail in \citetalias{Tarr:2017a} \S{3} using a ray--tracing WKB method; we briefly summarize that method in our \S{3}, below.  See especially \citetalias{Tarr:2017a} Figures 3-5, which show how different sections of an initial wave packet refract either toward or away from the null \change{due to local gradients in the phase speed of the wave, and their \S5.1-5.3, which discusses how the partition of wave energy changes for each section of the wave packet.}

\begin{figure}[ht]
  \includegraphics[width=0.45\textwidth]{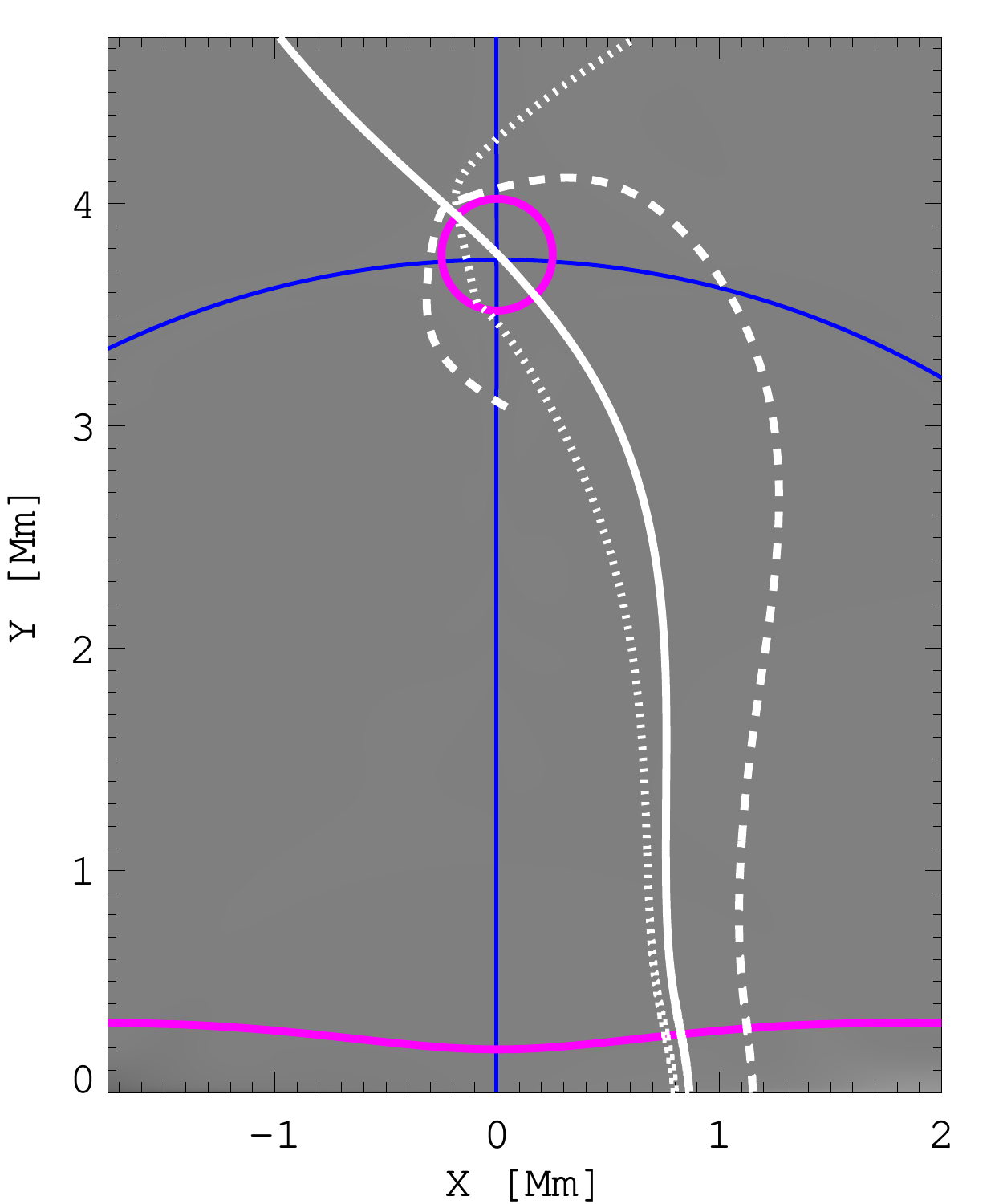}
  \caption{\label{fig:f2s} Three fast mode trajectories initialized along the lower boundary, corresponding to different portions of the wave packet.  The solid ray passes directly through the initial null location and leads to formation of the current sheet, the dotted ray reaches the equipartition region at an angle and converts to a slow ray.  The dashed ray first passes well to the right of the equipartition region, then reaches the mode conversion zones of the upper and left separatrices.}
\end{figure}

In the following work we focus only on the central wavepacket region that refracts toward the null: we will speak of the ``central portion'' of the pulse front and the ``wings,'' both referring to the just central region of the initial pulse, after the split.  \figref{fig:f2s} shows three example fast wave trajectories calculated as described in \S\ref{sec:formation} that illustrate the three basic ways that the central portion of the wave packet interacts with the null.

The initial pulse steepens as it propagates, and eventually gives rise to multiple shocks.  The plasma $\beta$ varies throughout the simulation, so the pulse shocks in distinct ways along different portions of the front.  The central portion propagates directly toward the null, approximately perpendicular to the magnetic field, and causes the null to collapse into a current sheet.  The solid line in \figref{fig:f2s} illustrates this path, while the energy densities are seen in \figref{fig:enfull}.

The wings of the pulse refract and wrap around the null, eventually reaching locations where they propagate nearly parallel to the field.  In \figref{fig:f2s}, this occurs where the dotted trajectory passes the equipartition contour.  At those locations, the plasma displacement vector and wave speeds are such that mode conversion is an efficient process \citep{Tracy:2003}.  \change{For an extensive discussion of the mode conversion process near the null and quantification of the energy exchanged between modes, see \citetalias{Tarr:2017a}, especially their \S5.}  In \figref{fig:enfull}, the mode conversion sequence can be seen by following the cyan regions near the equipartition contour at time $4.9t_N$, which convert to the magenta/white regions at $5.1t_N$, and finally the four strongest yellow regions at $5.3t_N$.  These latter are the primary converted slow shocks that propagate away from the null along the separatrices.  Behind each slow shock is a rarefaction wave followed by a reverse shock that brings the plasma nearly, but not quite, back to its original state.

Finally, at the far outer edges of the wave front (but still in the portion that refracts toward the null) the fast and slow wave speeds remain distinct so that the incoming fast waves do not immediately mode convert, but instead wrap around the null: see the dashed trajectory in \figref{fig:f2s}.  Eventually, these parts of the wings also reach locations where mode conversion is an efficient process, at which point they mode convert to generate secondary, weaker sets of slow shocks on the opposite sides of each separatrix: see, for example, the two yellow regions above and leftward of the current sheet at $t=5.3t_N$.

\begin{center}
  \begin{figure}[ht]
    \includegraphics[width=0.45\textwidth]{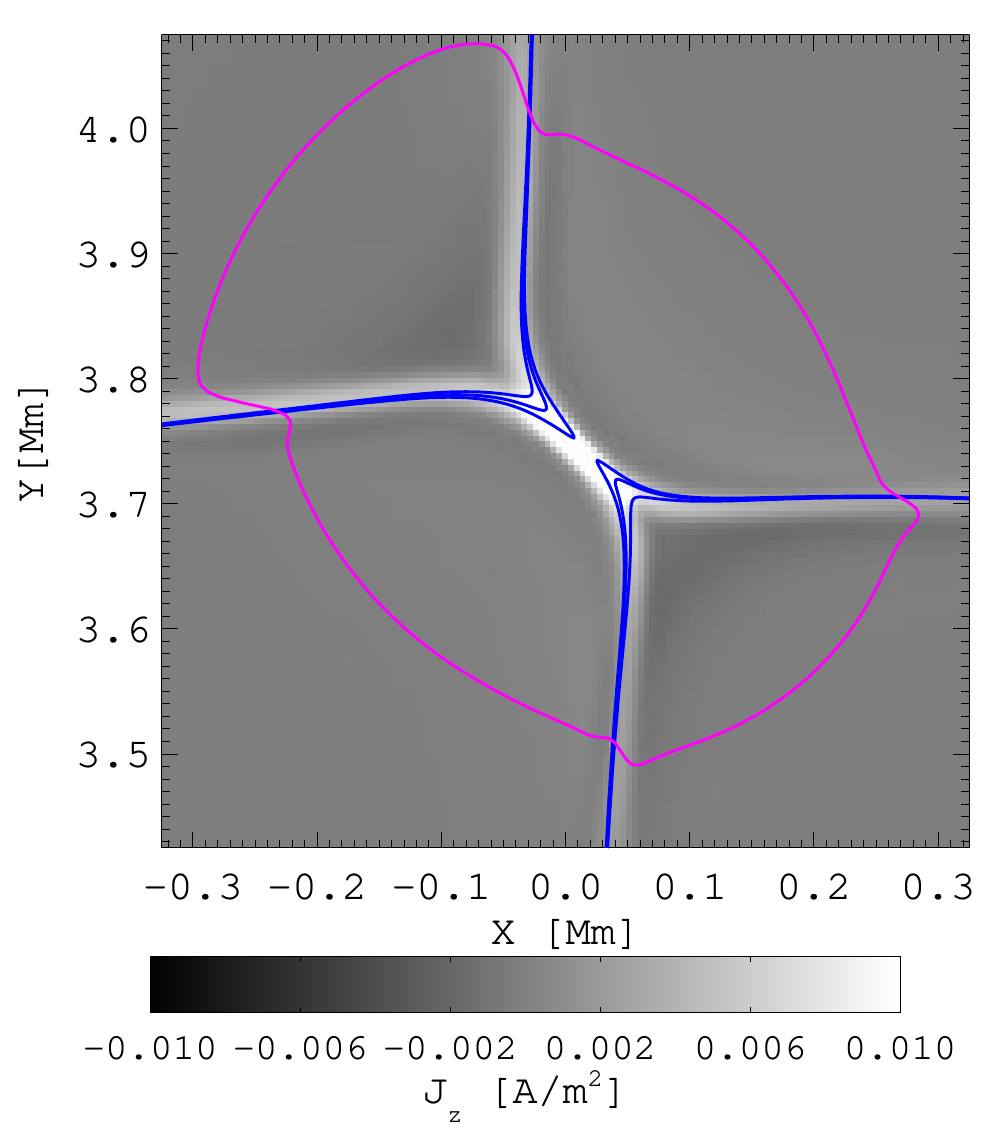}
    \caption{\label{fig:cs-cusp} Detail of current density and field line structure at time $t=5.625t_N$, around the maximum extent of the second current sheet that forms near the initial null location.  The blue curves are fields lines converging to the separatrices, and the equipartition contour is shown in magenta.}
  \end{figure}
\end{center}

Several more details are worth highlighting now.  As the central region of the pulse impinges on the null around $t=4.9t_N$, the central portion concentrates in the region where the magnetic field orientation is roughly perpendicular to the pulse front (\figref{fig:enfull}).  This fast pulse causes the null to collapse and form a current sheet oriented $+45^\circ$ counterclockwise (CCW) from the $\hat{\vect{x}}$ direction (\figref{fig:jzfull} at $5.3t_N$): the sheet is elongated parallel to the front.  The sheet then reconnects, shrinks, and eventually reforms in the $-45^\circ$ direction around $t=5.7t_N$, as shown in \figref{fig:cs-cusp}.  At that level of detail, we see that the separatrices connect to the current sheet in a cusp shape ($\prec$) instead of a Petschek ``Y'' shape, as described by \citet{Vekstein:1992, Vekstein:1993, Uzdensky:1997, Kulsrud:2011}.  The cusp shape is required by the presence of currents along the separatrices, which connect to the current sheet of the collapsed null.  This is true whenever the current sheet is present (e.g., also at $5.3t_N$, and during each later episode).

Subsequently, the sheet undergoes several more oscillations, damping all the while, as seen in \figref{fig:jzfull}.  These oscillations coincide with reconnection, a process that has been described in numerous other places, for instance, for a simulation of a flux tube emerging into a uniform vertical field \citep{Murray:2009}, and for a linear null point \citep{McLaughlin:2012a}.  We discuss the oscillatory reconnection further in \S\ref{sec:rx}.

Returning once more to the wings of the pulse, the mode--converted slow waves subsequently shock and the newly formed slow mode shocks propagate away from the null along each separatrix.  These shocks are distinct from the slow mode shocks of reconnection outflow, which do not propagate and instead remain attached to the collapsed--null current sheet.

The slow shocks lose energy as they propagate, heating the plasma in their wake.  Because a given shock that forms due to mode conversion is localized to one side of the separatrix, so is the heating.  As we show below, the shocks that form underneath the separatrix dome are stronger than those on the outside, so the heating is asymmetric across each separatrix.  The heated plasma expands to restore force balance, thus pushing the magnetic field away from a potential configuration and generating currents localized along each separatrix in the process.

The final state of the system therefore includes: (i) a tangential discontinuity across each separatrix that divides the more strongly shocked plasma on one side from the less shocked plasma on the other; (ii) an hourglass shaped region of increased internal energy density at the null related to reconnection outflows and dissipation within the collapsed--null current sheet; and (iii) persistent current density at the null and along each separatrix, as required to balance the new pressure gradients.  We detail each of these features in the following.
 
\section{Formation of Shocks and current sheets}\label{sec:formation}
In this section, we provide an analysis of the current densities which localize to the separatrix surfaces, as reported in \citetalias{Tarr:2017a}.  The formation and evolution of these current sheets are interwoven with the formation and propagation of various shocks in the simulation: the two studies cannot be decoupled.  A recent overview of the types of shocks that can arise in resistive MHD can be found in \citet{Goedbloed:2010} Chapter 20.

Since the (acoustically dominated) fast wave packet is injected away from $x=0$, as was the case in \citet{Tarr:2017a}, most of the initial wave energy efficiently transmits across the lower equipartition layer to become a magnetically dominated fast wave in the low $\beta$ portion of the simulation.  As described above, a section of the fast wave refracts toward the null.  Along the wave front's trajectory, the fast mode wave speed first increases upward from the photosphere but then decreases again near the null.  That latter decrease (even as the density drops) causes the wave to shock once it nears the null point; that is, it would not shock in the absence of a null, or at least a magnetic minimum.  Identifying the type of shock is difficult in this case: the pulse has some properties of a fast shock, but the transient dynamics of reconnection and mode conversion confuse the analysis.  We choose instead to analyze the forces near the null in the following.  On the other hand, the waves leaving the null behave like quasi--steady--state slow mode shocks.

To characterize the evolution of the shocks we start with the generalized Wentzel--Kramers--Brillouin \citep[WKB:][]{Weinberg:1962} approximation to find the trajectory of wave packets.  We present just the necessary definitions for the present analysis and refer the reader to \citetalias{Tarr:2017a} \S3.1 and Appendix A for details.  To lowest order, a shock will follow the same trajectory through an undisturbed background as a small--amplitude wave, but propagate at a faster speed \citep{Uralova:1994}.  Determining the trajectory amounts to integrating Hamilton's equations for a wave packet characterized by an initial location $\vect{x}(t=0)$ and wave vector $\vect{k}(t=0)$:
\begin{gather}
  \label{eq:hamil}\frac{d\vect{x}}{dt} = \vect{v}_g\text{ and } \frac{d\vect{k}}{dt} = -\vert k\vert \del v_{\phi}.
\end{gather}
Here, $\vect{v}_g=\partial_\vect{k}\omega$ is the group velocity, $v_\phi=\omega/\abs{k}$ is the phase speed, and $\omega$ is the frequency of the specific wave type: Alfv\'en, fast, or slow.  The phase and group velocities are calculated from the spatially--varying dispersion relation (see \citetalias{Tarr:2017a} \S3.1),
\begin{multline}
  \label{eq:dispersion-relation}\mathcal{D}(\omega,\vect{k},\vect{x})=0 = \Biggl(\omega_A^2- k^2v_A^2\cos^2\theta\Biggr) \times \\
  \Biggl(\omega_{\pm}^2 -k^2\Bigl[\frac{1}{2}(v_A^2+c_s^2)\pm\frac{1}{2}\sqrt{v_A^4+c_s^4 - 2v_A^2c_s^2\cos2\theta}\Bigr]\Biggr), 
\end{multline}
where $v_A$ and $c_s$ are the Alfv\'en and sound speeds, and $\omega_A,\ \omega_+$, and $\omega_-$ select the \change{Alfv\'en, fast, and slow branches of the dispersion relation, and each has a positive and negative solution, for a total of six solutions}.  We measure distance along a trajectory by $s$.  For each solution to $\mathcal{D}=0$, $\omega\propto k$ so the waves are dispersionless.  However, the waves refract as they propagate due to the spatial dependence of the phase and group velocities.  The magnitude of the wave vector divides out of Hamilton's equation, so hereafter we only refer to the normalized wave vector; e.g. the normal direction to a phase front has $\abs{\vect{k}}=1; \vect{k}=[\cos(\phi),\sin(\phi)]$ where $\phi$ is the angle measured counterclockwise from the positive $x$ direction.  Ray trajectories are found by numerically integrating equations \eqref{eq:hamil} using a Runge Kutta method.

\subsection{Current Sheet formation}\label{sec:CSformation}
\begin{center}
  \begin{figure}[ht]
    \includegraphics[width=0.45\textwidth]{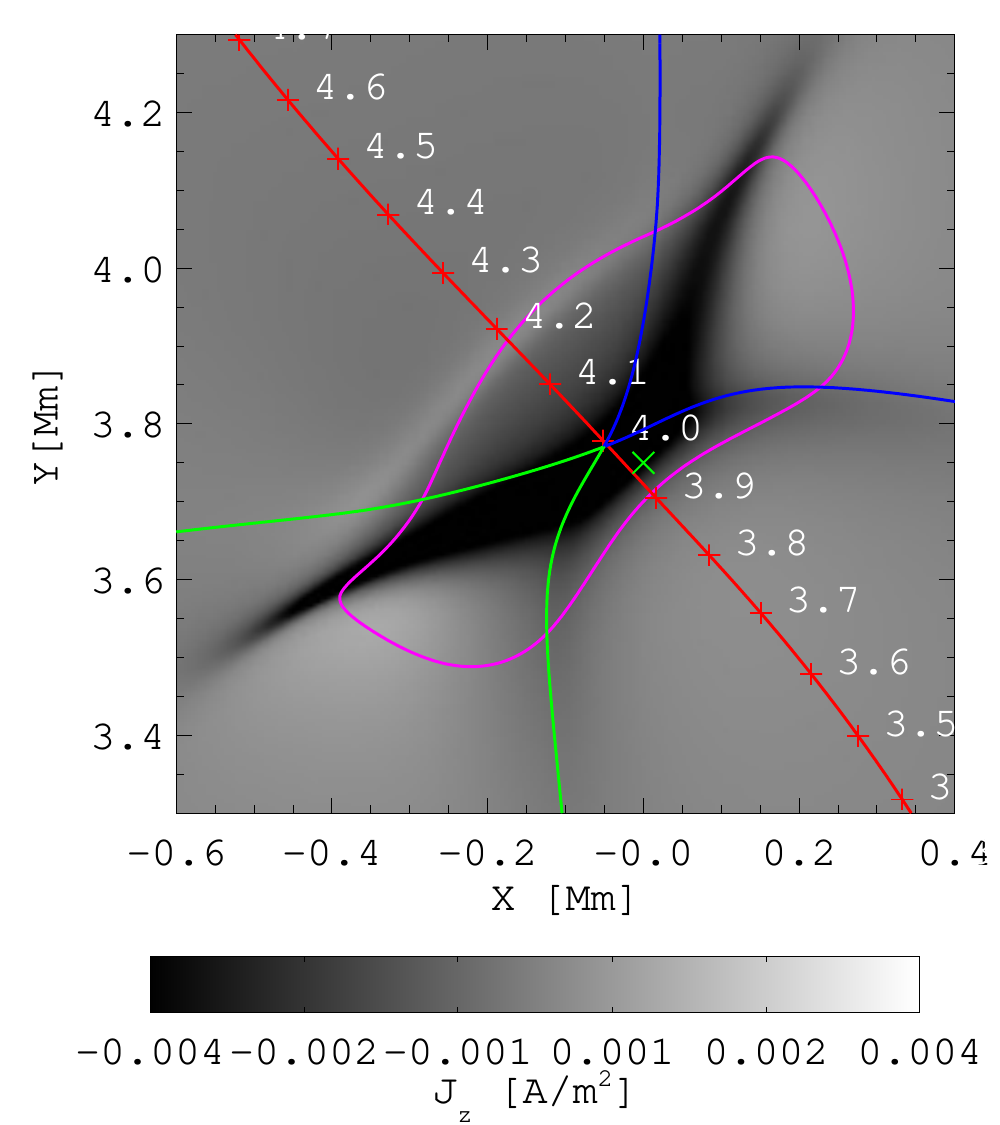}
    \caption{\label{fig:t203-jz} Contours of current density at time $t=4.95t_N$, as the first current sheet forms near the initial null location (green cross).  The red curve is the fast ray trajectory with distance from the initialization point indicated in $\unit{Mm}$ (see text).  Green and blue curves are field lines at the approximate location of the separatrix.  The (distorted) equipartition contour is shown in magenta.}
  \end{figure}
\end{center}

To study how the pulse causes the null to collapse into a current sheet, we first find a fast mode trajectory connecting the injected wave packet to the region around the null: Equations \eqref{eq:hamil} are solved for the $\omega_+$ branch of Equation \eqref{eq:dispersion-relation} with the initial conditions $\vect{x}(0)=(0.86, 0.005)\unit{Mm},\ \vect{k}(0)=(0,1)$; this trajectory (\figref{fig:multi-csva}), passes within a few computational cells of the initial null point at a distance $s=3.94\unit{Mm}$ along the curve.  The center of the initial current sheet is shown at time $t=4.95t_N$ in \figref{fig:t203-jz}.  Plasma parameters extracted along this curve for all output simulation times are analyzed below.  The long axis of the current sheet is perpendicular to the incoming fast ray trajectory, so the current sheet forms in the plane of the front, $\approx 45^\circ$ CCW from $+\hat{\vect{x}}$ \change{direction}.

\begin{figure}
  \includegraphics[width=0.45\textwidth]{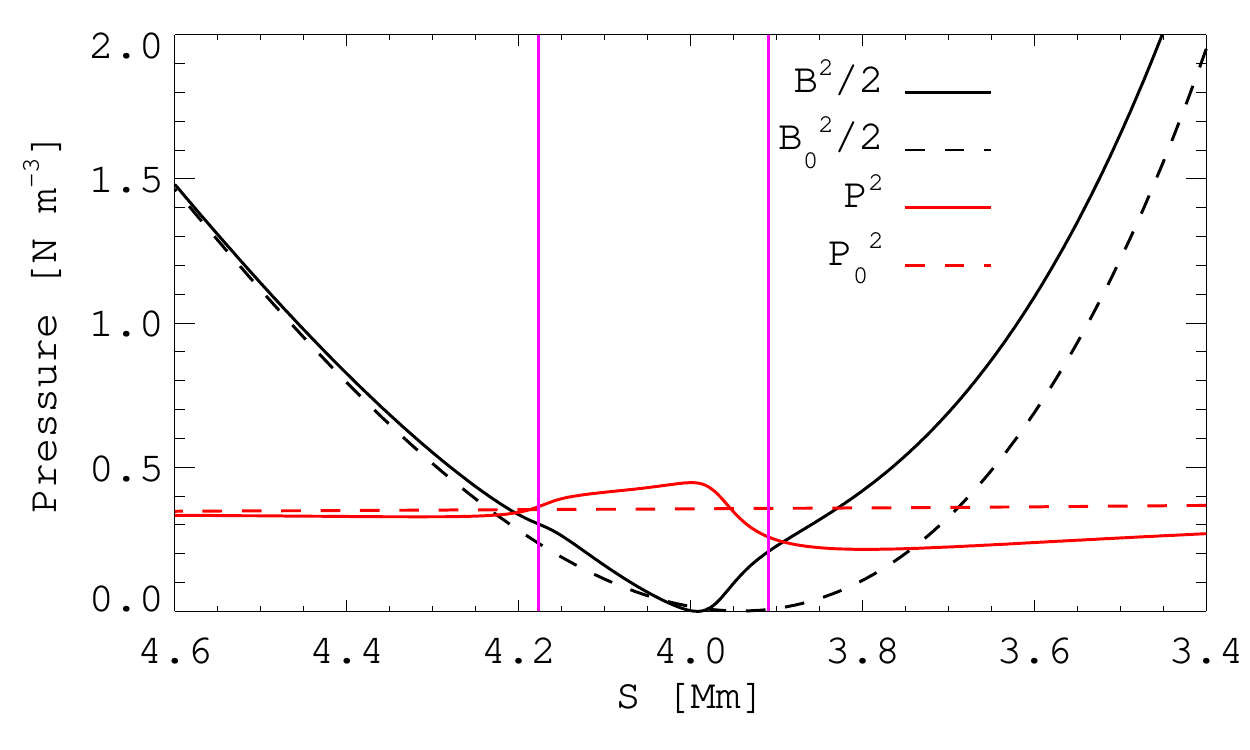}
  \caption{\label{fig:t203cs-props}  Magnetic pressure (black) and plasma pressure (red) extracted along the fast trajectory, at time $t = 4.95t_N$ (solid) and for the background state (dashed).  The vertical magenta lines mark the location of the distorted equipartition contour (where the magenta and red curves intersect in \figref{fig:t203-jz}).}
\end{figure}

\figref{fig:t203cs-props} shows the magnetic pressure $\frac{1}{2}B^2$ (black) and plasma pressure $P=(\gamma-1)\rho\epsilon$ (red) in the vicinity of the current sheet along the fast trajectory.  Solid curves are for time $4.95t_N$ and dashed curves for the background state.  The abscissa shows distance $s$ along the trajectory, oriented to agree with the direction of propagation in the simulation: if time were to progress, the pulse would move from right to left.  Derivatives of these curves are the magnetic and plasma pressure forces projected along the trajectory, which we discuss in depth below.

As the current sheet forms the magnetic field becomes more distorted and enhanced on the incoming side of the null (rightward in \figref{fig:t203cs-props}) compared to the outgoing side, indicating that the magnetic part of the pulse is not able to propagate across the null.  However, the pressure increases across the null, indicating that the pressure dominated part of the wave is, in contrast, able to propagate across the null.  From this analysis we found that the initial incoming fast wave is split into a magnetically dominated and an acoustically dominated pulse, each with separate dynamics.  The magnetic portion of the pulse has converted to a slow--type disturbance and is unable to cross the null, while the energy and density disturbance (the plasma pressure) transmits across the null as a fast wave, but carries substantially less energy.  This is consistent with the dynamics of wave splitting and propagation determined in \citetalias{Tarr:2017a} by numerically integrating the wave conservation relation \eqref{eq:wave-conservation} about the null.  The result, then, is that the current sheet forms due to a strong gradient in the magnetic pressure as flux piles up and is unable to rapidly leave the null region.

Because $\beta>1$ inside the equipartition region, we identify the pressure pulse that is able to cross the collapsing null as a transmitted fast wave, while we identify the magnetic pulse that stagnates and forms the current sheet as a slow wave.  The slow wave phase propagation vector $\vect{k}$ is perpendicular to the field, and so the slow phase velocity becomes 0.  But recall that the slow wave's \emph{group} velocity limits to the cusp velocity \change{$v_c = v_A c_s/\sqrt{v_A^2+c_s^2}$} directed along the magnetic field in this case \citep[][\S5.3.2]{Goedbloed:2004} so the newly formed slow wave packets propagate outward along field lines connected to the current sheet.  Meanwhile, the sheet's center moves along the fast mode trajectory as it forms, decelerating as it does so.  This is because the transverse forces on the sheet are unbalanced.  The sheet's length and maximum value of current density increase until $t\approx 5.04t_N$, after which the sheet begins to retract, weaken, and migrate backwards along the same trajectory, reconnecting all the while.

\begin{center}
  \begin{figure*}[ht!]
    \plottwo{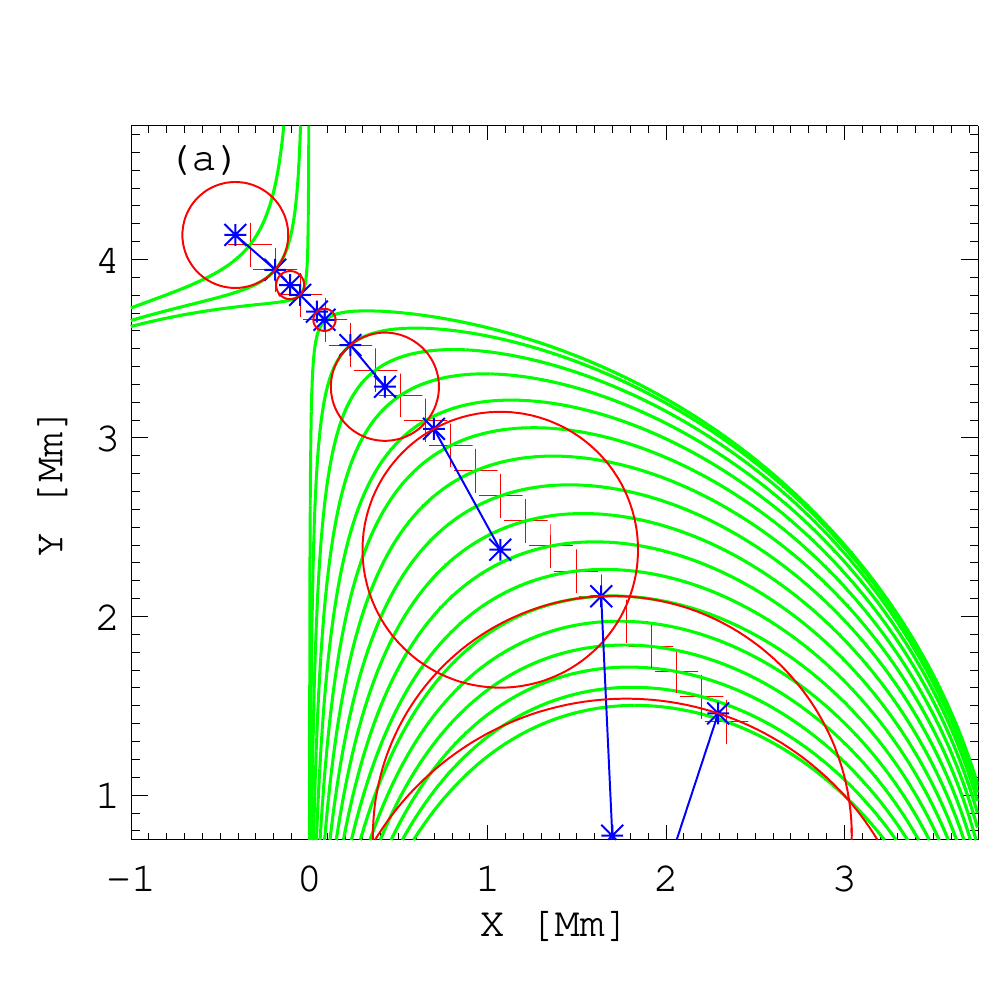}{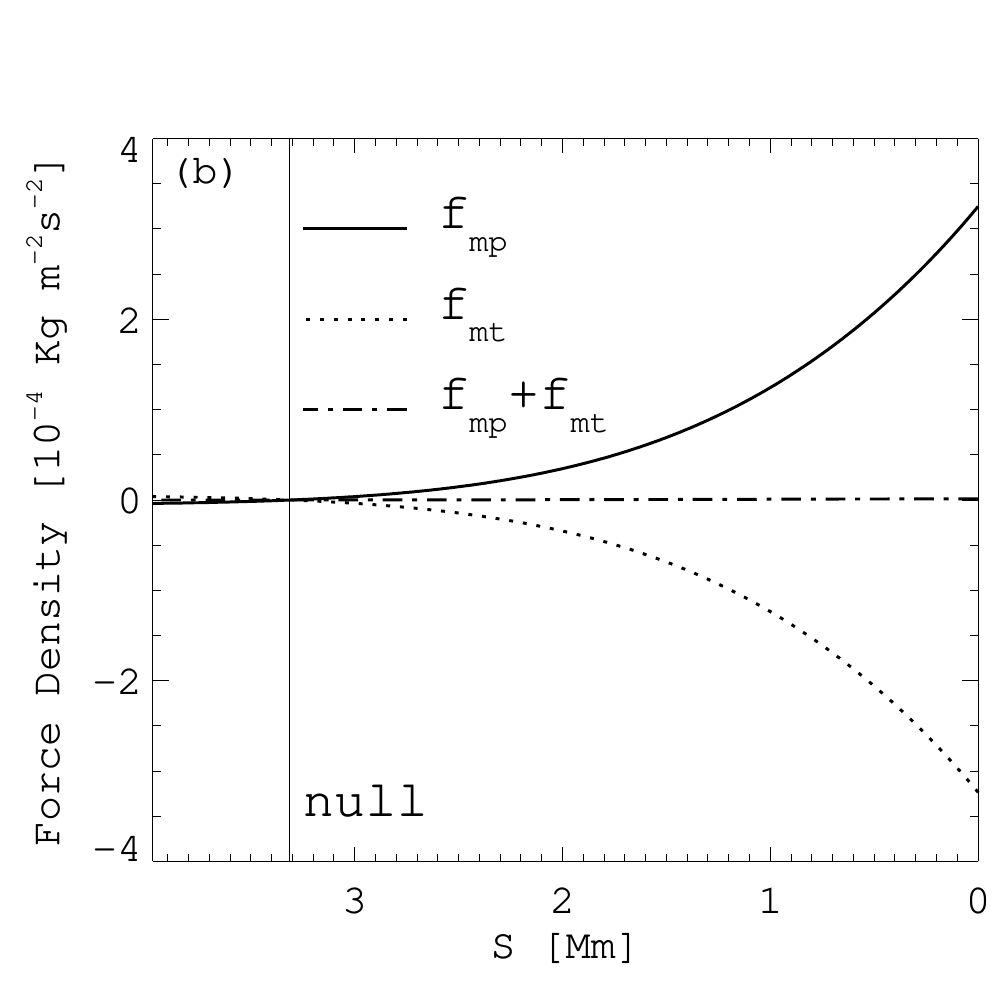}
    \caption{\label{fig:forcebalance-bk} (a): Magnetic field lines (green) and examples of the radius of curvature (red) and curvature magnitude and direction (blue), used in calculating the magnetic tension force; the blue lines connect the point at which the force is calculated to the center of curvature for that point.  (b): Magnetic pressure gradient ($f_{mp}$, solid) and tension ($f_{mt}$, dotted) terms in the Lorentz force density, and their sum (dash--dot), for the initial condition, calculated along the diagonal path between $\vect{x} = (2.3,1.4)$ and $(-0.47,4.2)\unit{Mm}$.  Positive values correspond force in the direction of increasing distance $s$.  The vertical line marks the location of the null.}
  \end{figure*}
\end{center}

We can get a better handle on the current sheet formation by considering the forces in the momentum equation\change{, written in the normalized units used in the code as}
\begin{equation}
  \rho\frac{d\vect{v}}{dt} = -\del P +\vect{j}\times\vect{B} + \rho\vect{g},
\end{equation}
where $\frac{d}{dt}=\partial_t + \vect{v}\cdot\del$ is the advective derivative.  Near the null, the gravitational force is several orders of magnitude smaller than the pressure and magnetic forces caused by gradients in background or the wave, so we will mostly ignore that term.  The Lorentz force $\vect{j}\times\vect{B}$ may be decomposed into two terms corresponding to a pressure and a tension,

\begin{align}
  \vect{j}\times\vect{B} & = (\del\times\vect{B})\times\vect{B} =  - (\del\vect{B})\cdot\vect{B} + (\vect{B}\cdot\del)\vect{B}\label{eq:fla}  \\
  & = -\del_\perp \frac{B^2}{2}+B^2(\hat{\vect{b}}\cdot\del\hat{\vect{b}})\label{eq:flb} \\
  & = -\del_\perp \frac{B^2}{2}+B^2\vect{t}\equiv \vect{f}_{mp}+\vect{f}_{mt}
\end{align}
where the curvature vector $\vect{t}$ defines the radius of curvature $R_c$ through $\abs{\vect{t}} = \frac{1}{R_c}$, and $\vect{B} = B\hat{\vect{b}}$.  Note that the parallel components of the derivative (along $\hat{\vect{b}}$) cancel between the two terms on the right side of \eqref{eq:fla}, so that the remaining terms are perpendicular to the field line: in the 2D case, they act in the same (or opposite) direction.

By design, the initial condition is in force balance.  \figref{fig:forcebalance-bk}(a) shows several examples of the curvature magnitude (red circles) and direction (blue lines), while \figref{fig:forcebalance-bk}(b) shows the magnetic pressure (solid) and tension (dotted) force density terms calculated separately, as well as their sum (dash--dot), along a $-45^\circ$ line passing through the null.  Near the null, the derivative of the plasma pressure along this line is initially nearly zero.  The derivative of the perpendicular magnetic pressure along the line smoothly transitions from negative below the null to positive above it, while the tension does the opposite.  Although $\beta>1$ around the null, the Lorentz force is still initially self--balanced.

\begin{center}
  \begin{figure*}[ht!]
    \plottwo{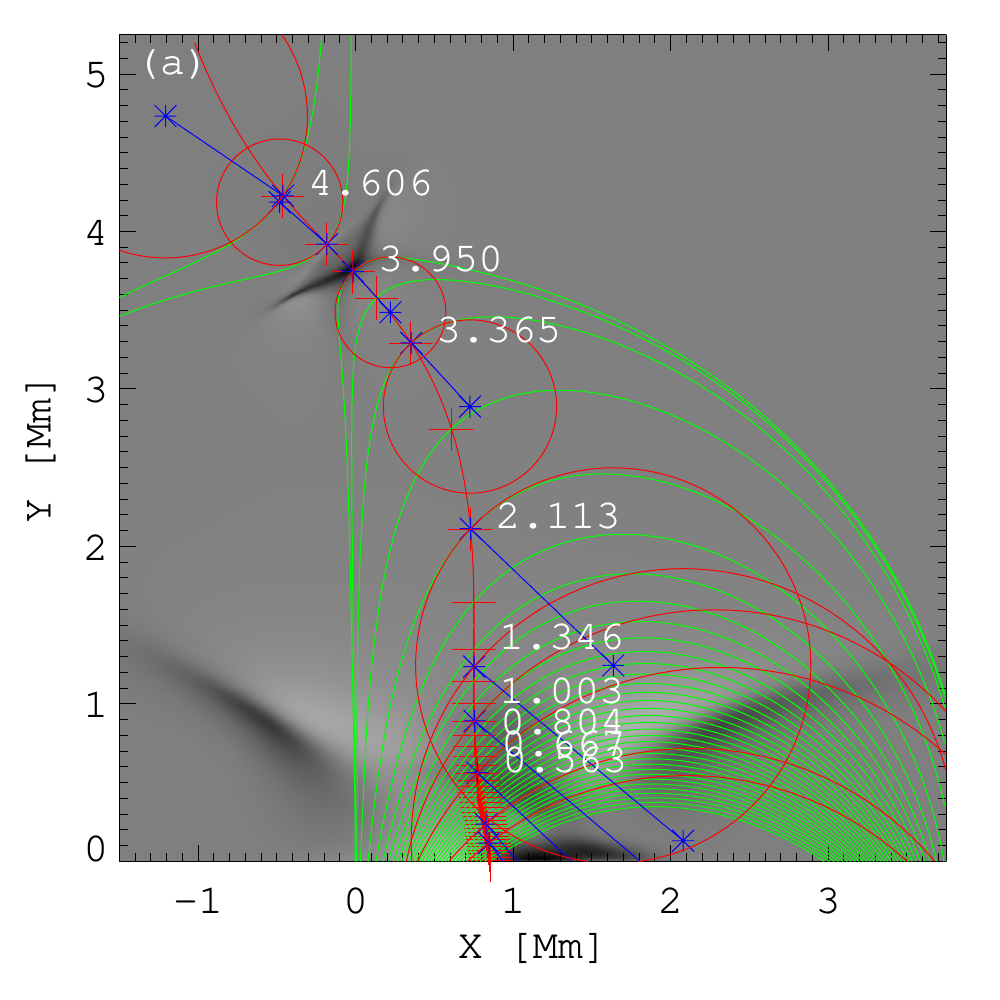}{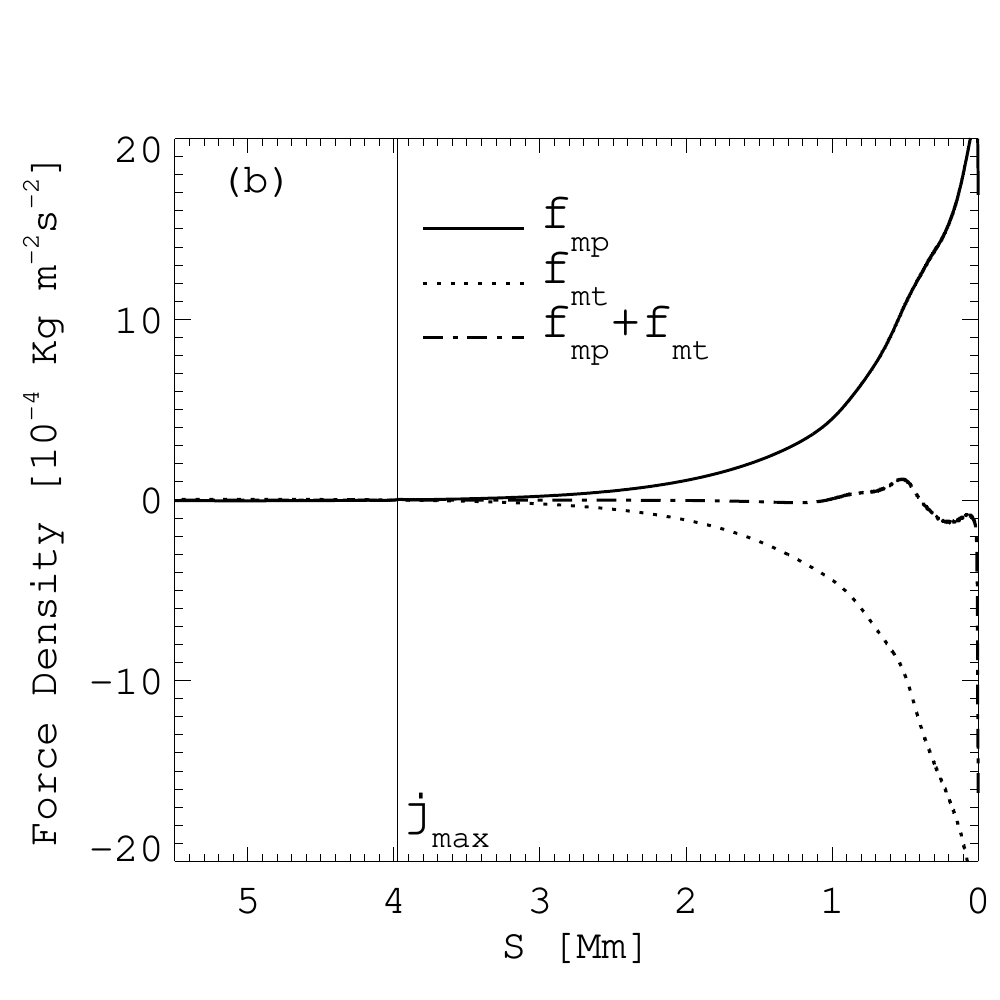}
    \caption{\label{fig:forcebalance-t203} Similar to \figref{fig:forcebalance-bk} for time $4.95t_N (120.2\unit{s})$.  Grayscale in (a) shows out--of--plane current density $j_z$.  Distances are indicated along the fast ray trajectory (red curve), and the magnetic forces calculated along that trajectory are shown in (b).}
  \end{figure*}
\end{center}

Figure \ref{fig:forcebalance-t203} shows similar force diagrams for time $4.95t_N$, where each force term is calculated along the solid fast ray trajectory from \figref{fig:f2s}, reproduced here as the red curve.  The grayscale in panel (a) is the current density $j_z$.  Panel (b) shows that the introduced wave pulse breaks the balance between the magnetic tension and magnetic pressure gradient forces, as easily seen at distances $<1\unit{Mm}$.

\begin{center}
  \begin{figure}[ht]
    \includegraphics[width=0.45\textwidth]{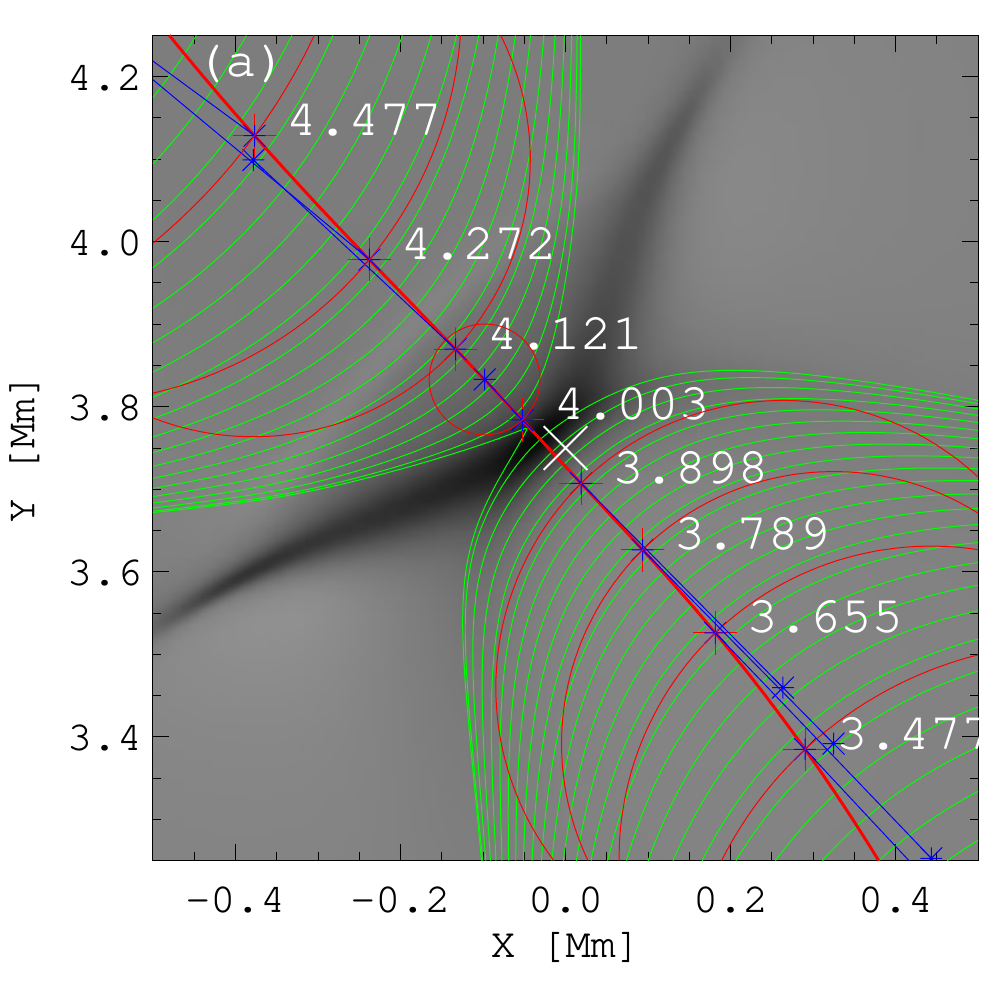}\\ \includegraphics[width=0.45\textwidth]{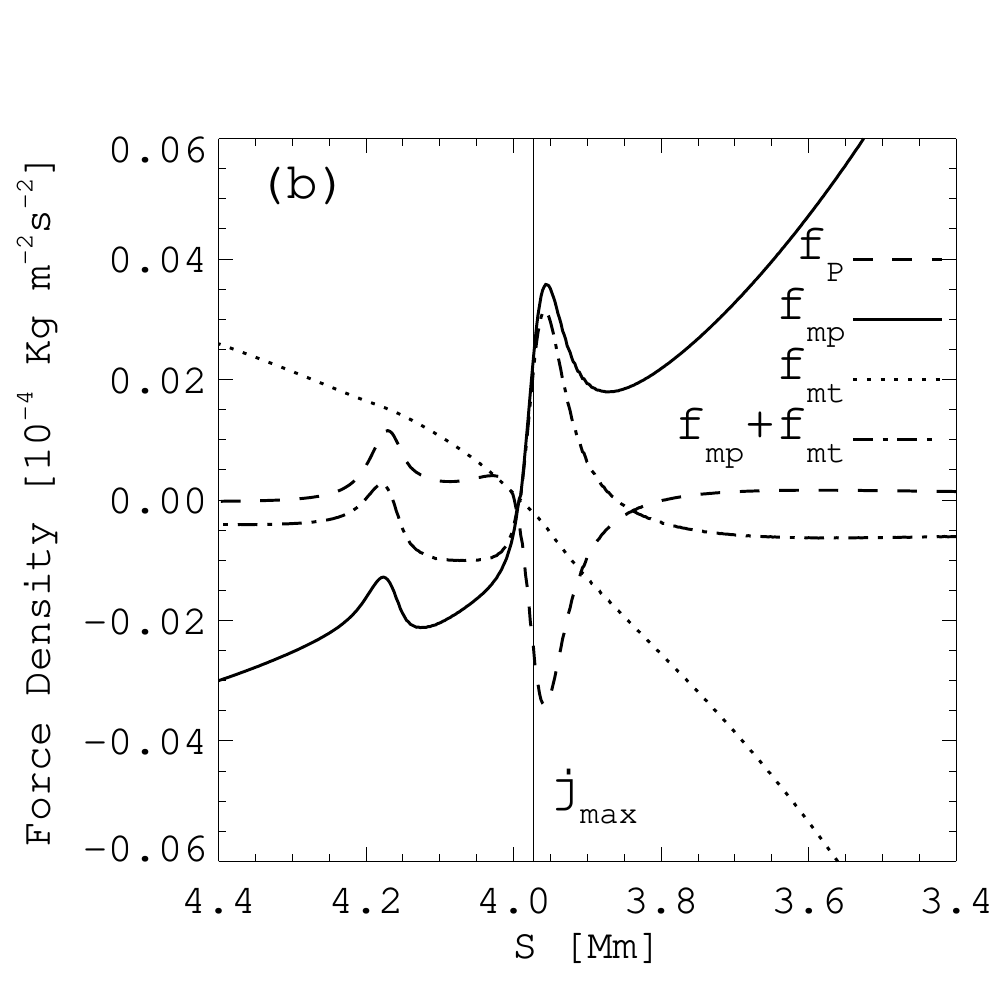}
    \caption{\label{fig:forcebalance-t203z} Same as \figref{fig:forcebalance-t203} for a zoomed in region around the nascent current sheet at time $4.95t_N$.  An animation of this figure is available in the online material.}
  \end{figure}
\end{center}

\begin{center}
  \begin{figure*}[ht]
    \includegraphics[width=\textwidth]{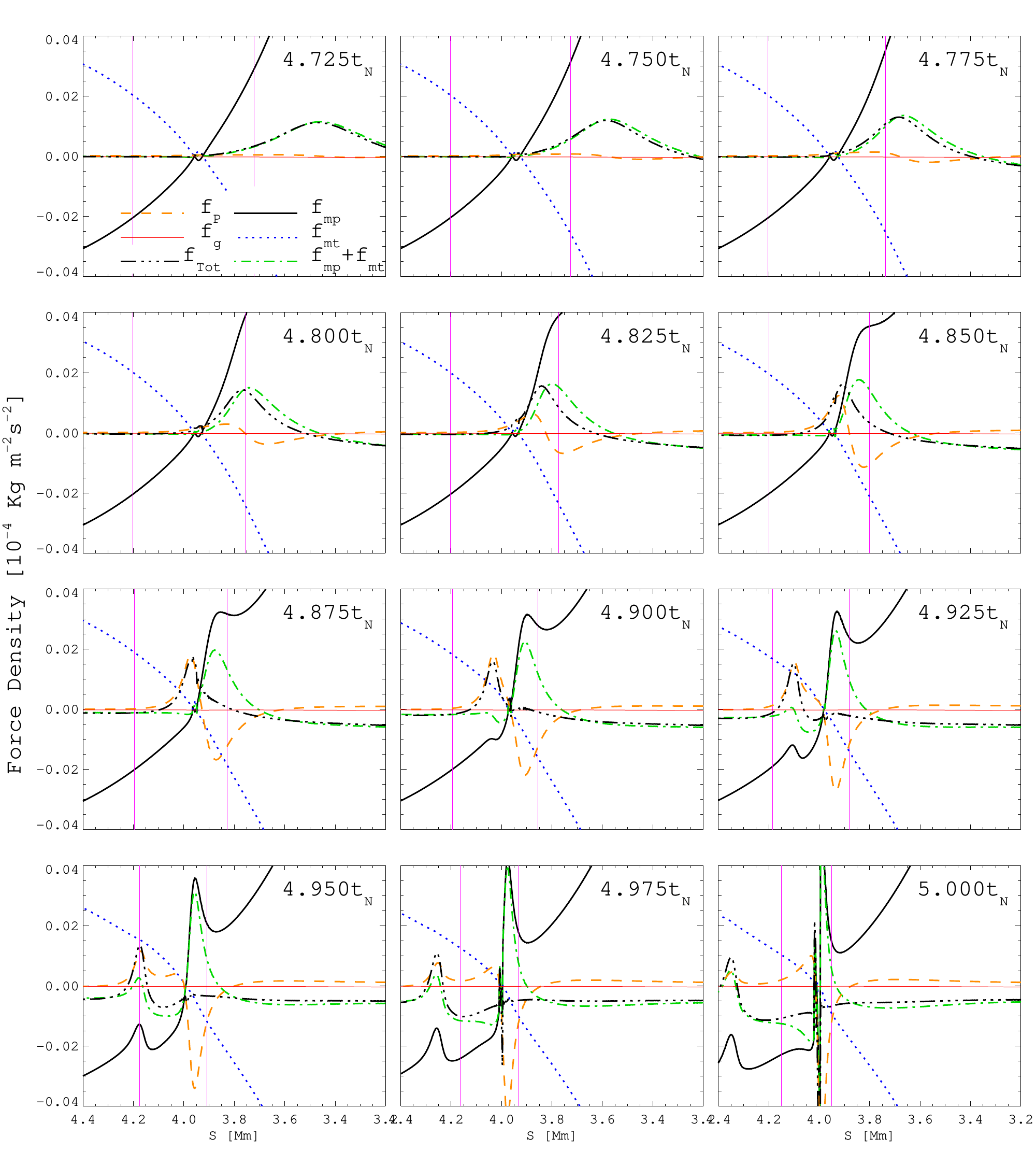}
    \caption{\label{fig:force-timeseries} Time series, where each panel is similar to \figref{fig:forcebalance-t203}(b), for a zoomed in region around the null.  An animation of this figure is available in the online material.  Each line is a force density term calculated along the fast ray, as labeled in the first panel: magnetic pressure gradient (solid black), magnetic tension (dotted blue), total Lorentz force (dashed-dot green), plasma pressure gradient (dashed orange), gravity (thin solid red, and essentially marks the $0$ line), and the total force density (black dash-dot-dot-dot).  Thin vertical magenta lines mark the equipartition surface.  The initial null point location is approximately at $3.84\unit{Mm}$.}
  \end{figure*}
\end{center}

A detail of \figref{fig:forcebalance-t203} near the null point is shown in \figref{fig:forcebalance-t203z}.  Panel (b) now also includes the plasma pressure gradient force (dashed).  A time series of the force balance near the null is shown in \figref{fig:force-timeseries} and the supplemental animation available in the online material.  The time series includes additional force terms and their combinations: the plasma pressure gradient force (dashed line), the gravitational force (solid red), and the total force (dash-dot-dot-dot).

Figure \ref{fig:forcebalance-t203z} highlights the balance of forces as the initial current sheet forms at time $4.95t_N$.  At this time, the initial pulse has split in two: there is the main pulse associated with the current sheet located at a distance $s=3.94\unit{Mm}$ along the trajectory, and a transmitted pulse around $s=4.2\unit{Mm}$.  The magnetic pressure gradient force and the plasma pressure gradient force are the derivatives of the black and red curves from \figref{fig:t203cs-props}, respectively.

Panel (b) shows that the tension force (dotted line) smoothly varies through zero across the collapsing current sheet, marked by the vertical line labeled $j_{max}$: the tension force does not cause the localization of currents.  Conversely, the magnetic pressure gradient force (solid) shows multiple local minima and maxima, with each extrema associated with extrema of the current density (compare to panel (a)).  The sum of the two magnetic terms, the total Lorentz force (dash--dot), shows that the magnetic pressure dominates the tension inside the pulses (e.g., at locations $\approx 3.9$ and $4.2\unit{Mm}$).  At the same time, a plasma pressure gradient (dashed) has built up to nearly counteract the magnetic pressure gradient.

As we now detail, the decoupling of the plasma pressure force from the magnetic pressure force at the null is what allows the current sheet to develop.  From the time series in \figref{fig:force-timeseries} we see that as the pulse approaches the null, the magnetic pressure gradient (solid) increases in magnitude relative to the background at the pulse's leading edge, and it decreases in magnitude behind the front.  This is most easily seen in the change in the solid line to the right of the null \change{$(s\approx3.8\unit{Mm})$} between times $4.825$ to $4.9t_N$.  During this time, the tension term (dotted) simply increases in magnitude everywhere behind front, though this is a subtle effect: it is most easily seen from the leftward drift over time of the dotted line along the lower boundary of each panel.

The total Lorentz force therefore has a localized leading force directed towards the null, due mostly to magnetic pressure, and a fairly uniform trailing restoring force away from the null, due mostly to tension.  This situation holds until a reflected pulse forms and the current sheet begins to retract at time $\approx5.1t_N$, as seen in the online animation.

The dynamics of the pulse rapidly changes once it crosses the equipartition region.  When the magnetic pressure pulse crosses the equipartition region, the plasma compression associated with it begins to cause an energetically important increase in plasma pressure: see the orange dashed line in \figref{fig:force-timeseries} starting at time $4.775t_N$.  The leading and trailing edges of the plasma compression give rise to pressure gradient forces oriented in the forward and backward directions along the fast trajectory (dashed line at $\approx 3.7\unit{Mm}, 4.8t_N$).  As the pulse reaches the location where the magnetic field switches direction (the null), two things happen at once: (i) the forward plasma pressure pulse is able to cross the null as a (still high--$\beta$) transmitted fast wave and becomes the dominant force on the other side of the null: this can be seen by comparing the total force (dash--dot-dot-dot) to just the magnetic force (dash--dot) at $t=4.9t_N$; (ii) the magnetic pressure pulse stalls on the incoming side of the null, and is mostly balanced by the trailing plasma pressure gradient pulse: compare the dashed and solid lines near $x=3.8\unit{Mm}$.

Once the transmitted fast wave crosses the equipartition region on the exiting side (leftward) of the null at $4.95t_N$, the plasma pressure gradient begins to lose its dominance while the magnetic pressure again starts to dominate the pulse dynamics: this pulse remains a transmitted fast wave\change{, now in the low--$\beta$ region}.  The online animation of this figure shows that this transition is complete by $5.05t_N$, when the transmitted fast wave is at $s=4.6\unit{Mm}$.  According to our calculations in \citetalias{Tarr:2017a}, the transmitted fast wave carries away $\approx 7\%$ of the energy to impinge upon the null.

\subsection{Reconnection and null point heating}\label{sec:rx}

The strong currents that localize to the null lead to both reconnection and Ohmic heating near the null.  Reconnection involves an extended area around the null where field lines are swept into the current sheet.  Because the field strength drops with height, these field lines map back to a narrow region at the lower boundary which straddles the separatrix near $x=0$.  To study the reconnection, we traced 1000 field lines from $y=0$ spanning $x=\pm 0.47\unit{km}$ (one tenth of a simulation voxel to either side of $x=0$) and tracked through time the locations of their opposite footpoints.  During reconnection, these jump discontinuously between the locations of the other two photospheric separatrices at $x=\pm3.75\unit{Mm}$.

\begin{figure}[ht]
  \includegraphics[width=0.45\textwidth]{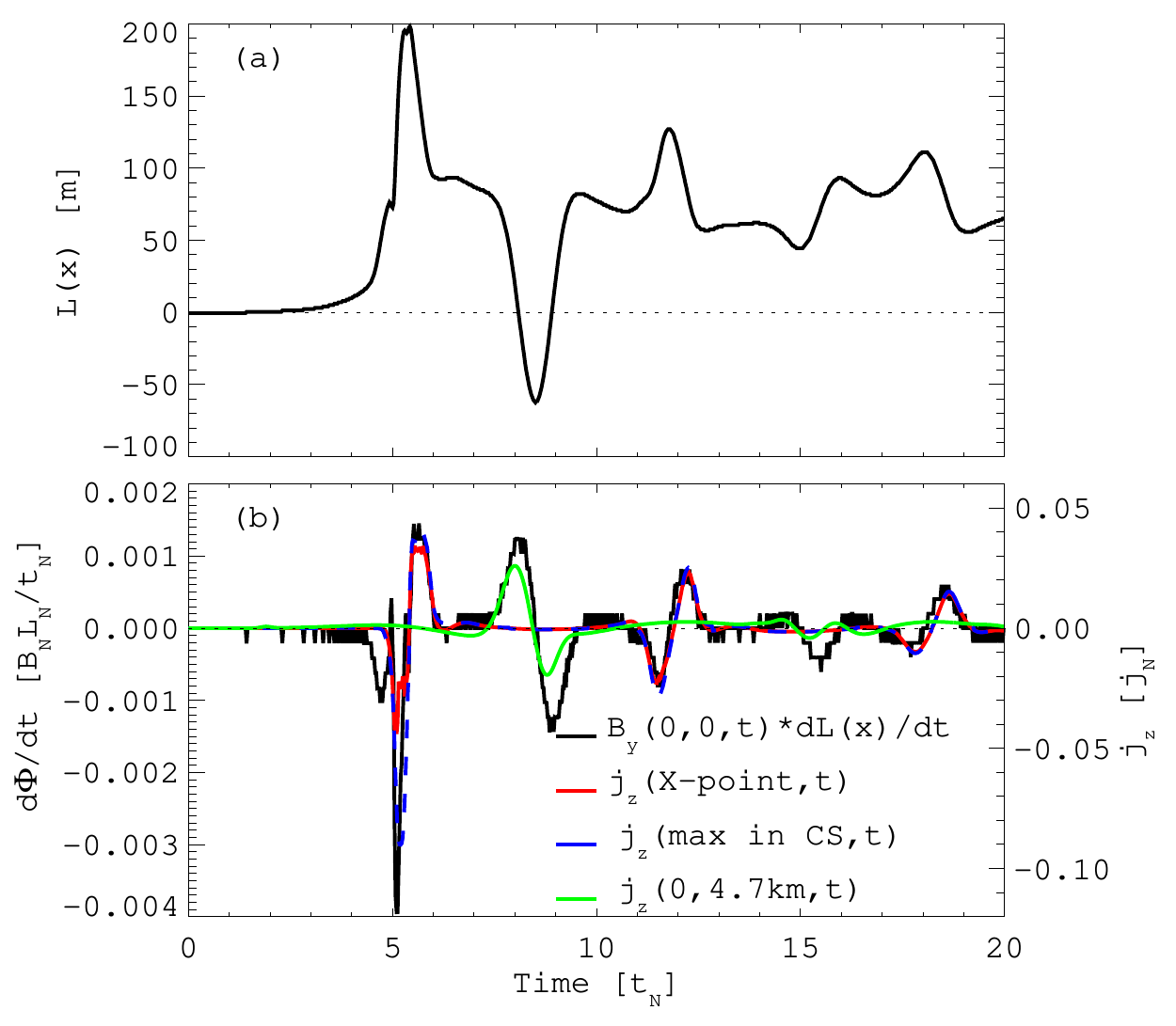}
  \caption{\label{fig:rxrate} (a) Location of central separatrix at $y=0$ over time as it shifts due to reconnection at the null.  (b) The reconnection rate, determined by integrating the flux swept out by the shifting separatrix location (left axis; black).  The out--of--plane current density $j_z$ is given on the right axis for several locations: at the \change{null} (red), the location of maximum current density in the current sheet (dashed blue), and just above the lower simulation boundary along the $x=0$ separatrix (green).   The scaling between the left and right axes is given by $\dot{\Phi}\approx \eta j_z$, where $\eta$ is the uniform simulation resistivity.}
\end{figure}

\figref{fig:rxrate} describes the reconnection over the course of the simulation.  Panel (a) shows $L(x)$, the position of the central separatrix along the lower boundary.  This separatrix moves laterally as reconnection transfers field lines through the current sheet.  Several reconnection episodes are evident, but the secular trend is to reconnect field lines between the right domain inside the dome and the left domain outside the dome, which shifts the separatrix in the \change{$+\hat{\vect{x}}$} direction.  This is consistent with reconnection across the initial current sheet.  Note that the lower boundary is line--tied with $v_x(x,0) = 0$, so the shift of the separatrix is due to a change of topology.

The rate at which flux $\Phi$ is swept out as the separatrix location changes with time gives the reconnection rate $\dot{\Phi}$, which is shown in black in \figref{fig:rxrate}(b); the finite number of traced field lines cause the discrete steps.  The reconnection rate should be related to the current density at the \change{null}, where reconnection occurs, by $\dot{\Phi} = \eta j_z$, and $\eta = 0.0333$ is the uniform simulation resistivity.  We determine the location of the null by finding the minimum of $B^2$ using a Newton--Raphson method, and then perform a bilinear interpolation of the current density $j_z$ to that point from the closest grid locations.

The red curve in \figref{fig:rxrate}(b) shows the \change{null} current density as a function of time, given in the normalized units $j_N$ on the right axis.  The blue and green curves show the maximum amplitude current density in the current sheet and the current density at $y=4.7\unit{km}$, one gridpoint above the lower boundary along the $x=0$ separatrix.  The null location is typically (but not always) the location of maximum magnitude current density in the current sheet.  The scaling between the right and left axes is indeed $\eta$.  Comparing the current density to the reconnection rate, we see that there appear to be two distinct processes at work: (i) one where the current density at the null scales with the reconnection rate like $\eta$ (e.g., times $\approx 5t_N,\ 12t_N, 18t_N$), and a (ii) second process where the reconnection rate scales (rather roughly) with the current density near the lower boundary ($\approx8t_N,\ 15t_N$).  The first set appear to be genuine reconnection at the null point.  The second set appear to be diffusion caused by strong gradients in the current density near the lower boundary just as the fast wave reflects off that boundary.  The side and top boundaries are far removed from the null so that, combined with the damping terms near these simulation boundaries, no reflections from these locations impinge upon the null.

The strong currents at the null in process (i) occur every time a fast wavefront sweeps across the null.  The first of these ($5t_N$) is due to the initial wavepacket reaching the null.  Note from \figref{fig:rxrate} that this episode has two reconnection phases, where reconnection happens first in one direction (peak at $5t_N$) and is immediately followed by reconnection in the opposite sense (peak at $5.5t_N$).  There does not appear to be further oscillatory reconnection associated with this episode.  This is also the case for the other two process (i) episodes at $12t_N$ and $18t_N$.  These latter two episodes correspond to the fast wave traveling back from the null to the lower boundary, reflecting, then finally refracting back to the null point.  Again, they do not seem to be a form of oscillatory reconnection inherent to the dynamics of the null.  Each time process (i) occurs it generates a substantial reconnection episode, influencing a large area around the null.  This implies that the processes of fast wave refraction into the null region, flux pile--up, and mode conversion are important for Ohmic heating within the null region.

In contrast, process (ii) does not coincide with strong currents at or around the null point, meaning its contribution to null point heating is minimal.  Instead, it appears to be due to diffusion of magnetic field caused by strong currents near the lower boundary, cotemporal with the fast wave reflection off the lower boundary.  It is an integral effect over the separatrix field line, leading to the lesser correlation between the green and black curves in \figref{fig:rxrate}(b) compared to the red, blue, and black curves.  Its amplitude also diminishes rapidly, as less of the fast wave energy returns to the lower central separatrix.  The similarity in the amplitudes between process (i) and process (ii) is due to the uniform resistivity, which was set to just exceed the numerical resistivity for this simulation.

\begin{center}
  \begin{figure}[ht]
    \includegraphics[width=0.45\textwidth]{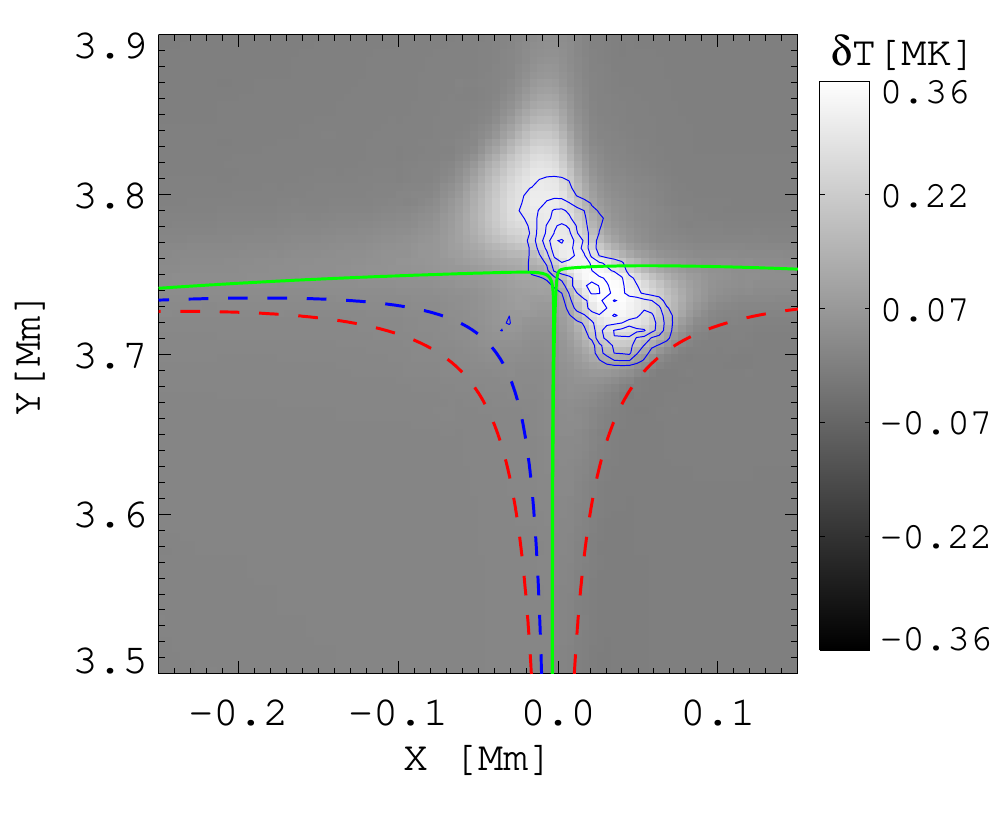}
    \caption{\label{fig:en-dumbbell} Grayscale: temperature difference at time $t=20t_N$ relative to the background stratification.  Blue contours show the approximate distribution of Ohmic heating, described in the text.  Field lines that bound the flux involved in reconnection are shown as dashed red lines, the field line associated with the initial separatrix is the dashed blue line, while the final separatrix is the solid green line.}
  \end{figure}
\end{center}

The grayscale in \figref{fig:en-dumbbell} depicts the change in temperature near the null point at the final simulation time, $t=20t_N$, relative to the baseline simulation where no wave packet was injected.  A dumbbell shaped region of enhanced temperature is located near the null, $\approx0.3\unit{MK}$ or $40\%$ above the temperature of the background stratified atmosphere.  The plasma temperature along each separatrix is enhanced by $\approx3-5\%$.  The separatrix dome is highlighted in green, and red dashed curves mark the extent of field lines underneath the dome that were involved in reconnection, as determined in \figref{fig:rxrate}(a).  The field line associated with the initial separatrix surface is shown as the dashed blue line and demonstrates the direction of flux transfer.  The plasma on opposite sides of the initial current sheet has been heated, and this heating is bounded by the field lines involved in reconnection.  The distribution of plasma density shows essentially the same features as temperature distribution, but in the opposite sense: the final density is reduced by $\approx 30\%$ relative to the background within the dumbbell region, and by $\approx 1-2\%$ along the separatrices.

To further determine if the reconnection is associated with heating around the null we integrated the Ohmic heating term $H_O = \eta j^2$ in time for a set of traced plasma elements.  Because we did not include conduction in our model, the heating integrates with each fluid element over time.  We therefore calculated the integrated displacement of a $(51\times51)$ grid of points surrounding the initial location of the null: $(-0.094<x< 0.140, 3.66<y<3.89)\unit{Mm}$.  The location of each of point as a function of time is given by $\vect{r}(t)=\vect{r}(0)+\int_0^t\vect{v}(t^\prime)dt^\prime$.  This method allows only a rough comparison because we used the simulation output at a cadence of \change{$dt\approx150\delta t$}, where $\delta t$ is the fundamental simulation timestep given by the Courant condition.  At each time we found the current density at the displaced locations by bilinear interpolation from the closest Eulerian grid points.  We next mapped the total Ohmic heating $\int H_O(t)dt$ at each Lagrangian point over time back onto the Eulerian simulation grid at the final time, and compared the integrated Ohmic heating with the final temperature of the plasma.  Contours of the Ohmic heating are displayed as solid blue lines in \figref{fig:en-dumbbell}.  The temperature and Ohmic heating distributions do overlap, which indicates that Ohmic heating within the current sheet is responsible for a stable, dumbbell--shaped high energy region around the null at late times.

Interestingly, the dumbbell shaped enhanced energy region is oriented perpendicular to the plane of the original current sheet, which forms parallel to the incoming wave front.  Comparison between the Lagrangian traces, the plasma density, and the current density reveals why.  Reconnection outflows from the ends of first the current sheet cause a mass density depletion within the sheet.  Most of the heating takes place inside the first current sheet while it is at its maximum extent, but this happens after the density depletion.  Ohmic heating is thus able to heat the remaining plasma to high temperatures so that the overall pressure near the null is enhanced relative to the initial state.  As the high pressure, high magnetic energy current sheet collapses and begins to reconnect in the opposite sense, the previous strongly heated regions get pushed out along the original inflow directions.  The result is the high temperature, low density, dumbbell shaped region oriented perpendicular to the original wave front.

After the first reconnection episode ends around $6.0t_N$, the plasma density remains low and the temperature high relative to the initial state.  In each later reconnection episode, the plasma temperature continues to increase by Ohmic heating so the pressure becomes locally enhanced and has strong gradients.  In between each episode the plasma relaxes until the pressure has minimal gradients.

\subsection{Slow shocks along the separatrices}\label{sec:shocks}
In this section we study the mode--converted slow waves that form along each separatrix.  As we will see, these waves eventually form slow mode shocks.  We study the area around the left separatrix in detail, but the same analysis holds around all four separatrices.  Note that, in particular, it holds around the upward separatrix that funnels wave energy higher into the corona.  This brings to mind the observations of spicules, which we discuss briefly in \S\ref{sec:discussion}.

The dashed curve in \figref{fig:multi-csva} labeled ``Slow Ray'' shows a solution to Equations \eqref{eq:hamil} for the slow mode branch of Equation \eqref{eq:dispersion-relation} with initial conditions $\vect{x}(0)=(-0.563, 3.436)\unit{Mm},\ \vect{k}(0)=(0,1)$: it traces out the trajectory of a mode--converted slow mode wave.  The initial condition was selected to describe the leading edge of the wave pulse in the mode conversion region around the time when the initial pulse converts, at $t=4.925t_N$.  The intersection of the wave front with the region of substantial mode conversion creates an extended region where the slow shocks form, all propagating on similar trajectories, essentially creating a shock tube.  The case we analyze is roughly in the middle of this tube.

The shock itself travels faster than a slow wave (it is a shock) but follows the same trajectory.  \citet{Afanasyev:2012} discuss a ray tracing method that captures this behavior, where the only essential difference is the replacement of the phase velocity $a$ by $(a+\frac{1}{2}\kappa U_{sh})$ in the $\hat{\vect{k}}$ term of the ray equation, where $U_{sh}$ is the shock velocity, $\kappa$ is a numerical coefficient dependent on local plasma properties and propagation angle to the magnetic field, and the shock velocity depends on the shape of the waveform that generated the shock.  We do not perform that analysis here, but simply determine the slow trajectory and then extract all simulation output along that trajectory at every snapshot in time.

We have performed the extraction along many trajectories, and as expected the plasma properties do not vary significantly for nearby trajectories within the shock tube.  The shock strength varies from path to path, decreasing away from the separatrix.  Inspecting the various plasma parameters along the slow trajectory from \figref{fig:multi-csva} as a function of time shows that the shock fully forms after traversing \change{$\approx1\unit{Mm}$} along this path: the $*$ symbols in \figref{fig:multi-csva} are spaced at $1\unit{Mm}$ intervals.

\begin{figure}[ht]
  \includegraphics[width=0.45\textwidth]{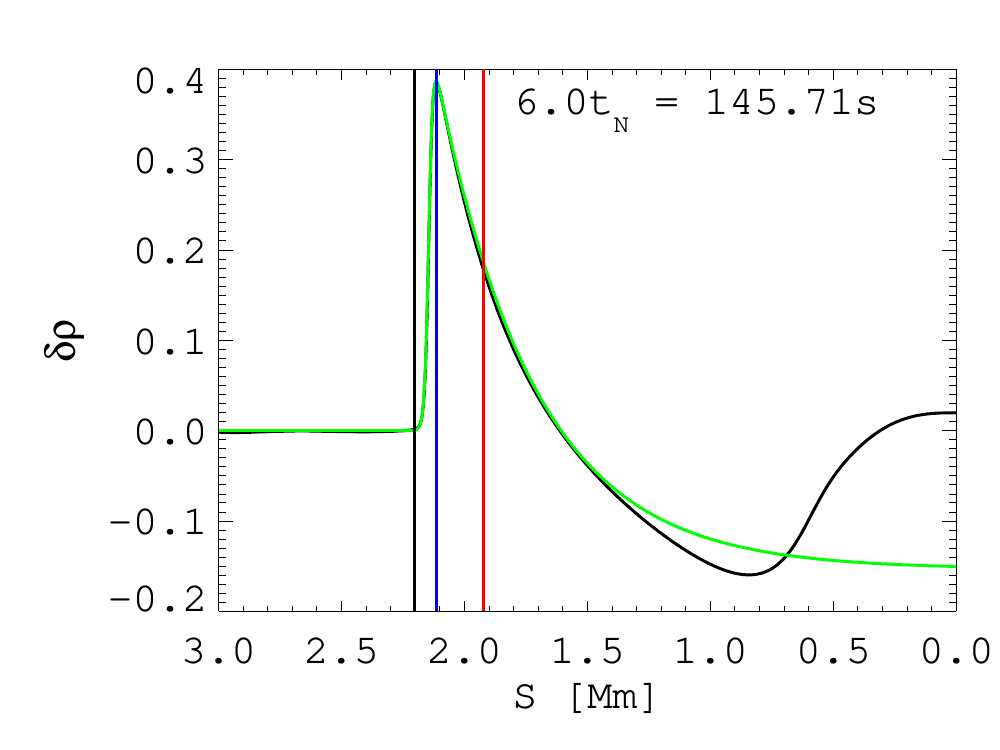}
  \caption{\label{fig:sk-still} Fractional density change along the slow shock trajectory at time $t=6.0t_N$.  The vertical lines mark the preshock and postshock (shock peak) locations (black, blue),  as well as the location an infinitesimal slow wave would have traveled starting at $s=1\unit{Mm}$ at time $4.925t_N$, when the fully formed shock peak passed that location (red; see text).  The green line is a fit to the shock front.}
\end{figure}

To define important features of the shock we introduce the fractional density change along the shock path relative to the background,
\begin{equation}\label{eq:drho}
  \delta\rho(s;t)=\frac{\rho(s;t)-\rho_0(s)}{\rho_0(s)},
\end{equation}
where $s$ is the distance along the shock trajectory.  \figref{fig:sk-still} shows an example of $\delta\rho$ at time $t=6.0t_N$.  The form of the wave pulse resembles a standard N--wave, with a forward shock front, rarefaction wave, and a reverse shock at the tail \citep{Courant:1948}.  Both the rarefaction wave and the tail, however, display more complicated behavior than for the strictly hydrodynamic N--wave \citep{Ro:2017}, pointing to \change{the} unique behavior of MHD shocks and the possible influence of reconnection outflows on the tail.

In order to automatically and robustly track the shock evolution we require a consistent definition of the preshock value and the shock peak.  To do so, we fit the shock front with a combination of a hyperbolic tangent and an exponential decay behind the peak.  The resulting functional form is
 \begin{equation}
   S(s) =\frac{1}{2}\Bigl(\tanh\frac{s_c-s}{s_w} + 1\Bigr)\Bigl(\exp(a + \change{bs}) + c\Bigr).
 \end{equation}
 The coefficients $\{s_c, s_w, a, b, c\}$ are determined using \fnc{lmfit.pro}, IDL's built--in implementation of the Levenberg--Marquardt algorithm.  The first two coefficients set the location and width of the shock front, while the latter three set the shock strength, decay rate, and depth of the rarefaction wave behind the shock, respectively.  We define the preshock location as the point where the normalized $\tanh$ function reaches $0.001$ of its maximum and the postshock location (the shock peak) as the zero of the second derivative of the fitted profile; the former is calculated directly from the analytic function while the latter is determined numerically using a Newton--Raphson method.

The shock's fractional density profile (black), analytic fit to the shock front (green), and locations of the preshock (black vertical) and postshock (blue vertical) locations are shown in \figref{fig:sk-still} at time $t=6.0t_N$.  For reference, the red vertical line in \figref{fig:sk-still} demarks the location where the shock peak would be if it traveled at the slow mode speed after fully forming at $s=1\unit{Mm}$ at time $t=4.925t_N$, when the fully formed shock peak passed that location.

\begin{figure}
  \includegraphics[width=0.45\textwidth]{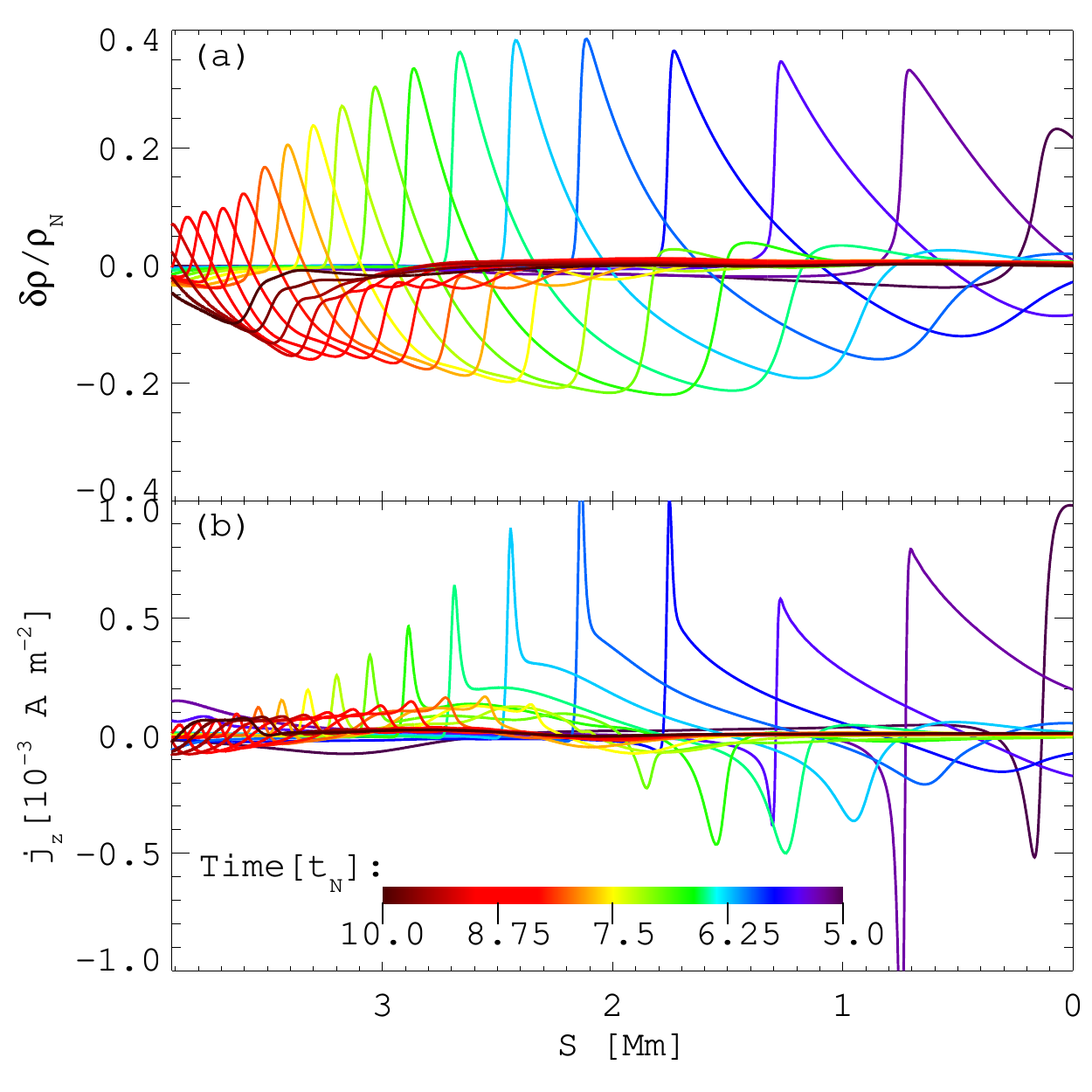}
  \caption{\label{fig:slow-shock-timeseries} Temporal evolution of $\delta\rho$ and $j_z$ along the slow shock trajectory from \figref{fig:multi-csva}. The color bar indicates time.}
\end{figure}

\figref{fig:slow-shock-timeseries} shows the temporal evolution of the fractional density (top) and current density (bottom) profiles along the slow shock trajectory; the color scale indicates time.  The amplitude of the shock increases until $t\approx 6t_N$ due to the convergence of the surrounding slow mode trajectories, which are guided by the convergence of field lines away from the null.  After that time, even though the field lines and slow trajectories continue to converge, shock dissipation becomes the dominant effect, and the shock amplitude decreases.  As the shock dissipates and its amplitude decreases, so does its propagation speed.  After $t=6.5t_N$ it essentially travels at the slow wave speed (the distance between the red and blue vertical lines in the time evolution of \figref{fig:sk-still} remains fixed).  A sharp spike in current density forms at the leading edge of the shock (see \figref{fig:slow-shock-timeseries}(b) at \change{$t\approx6t_N,\ s\approx1.75\unit{Mm}$}), whose origin is currently unexplained.

\begin{center}
  \begin{figure}
    \includegraphics[width=0.45\textwidth]{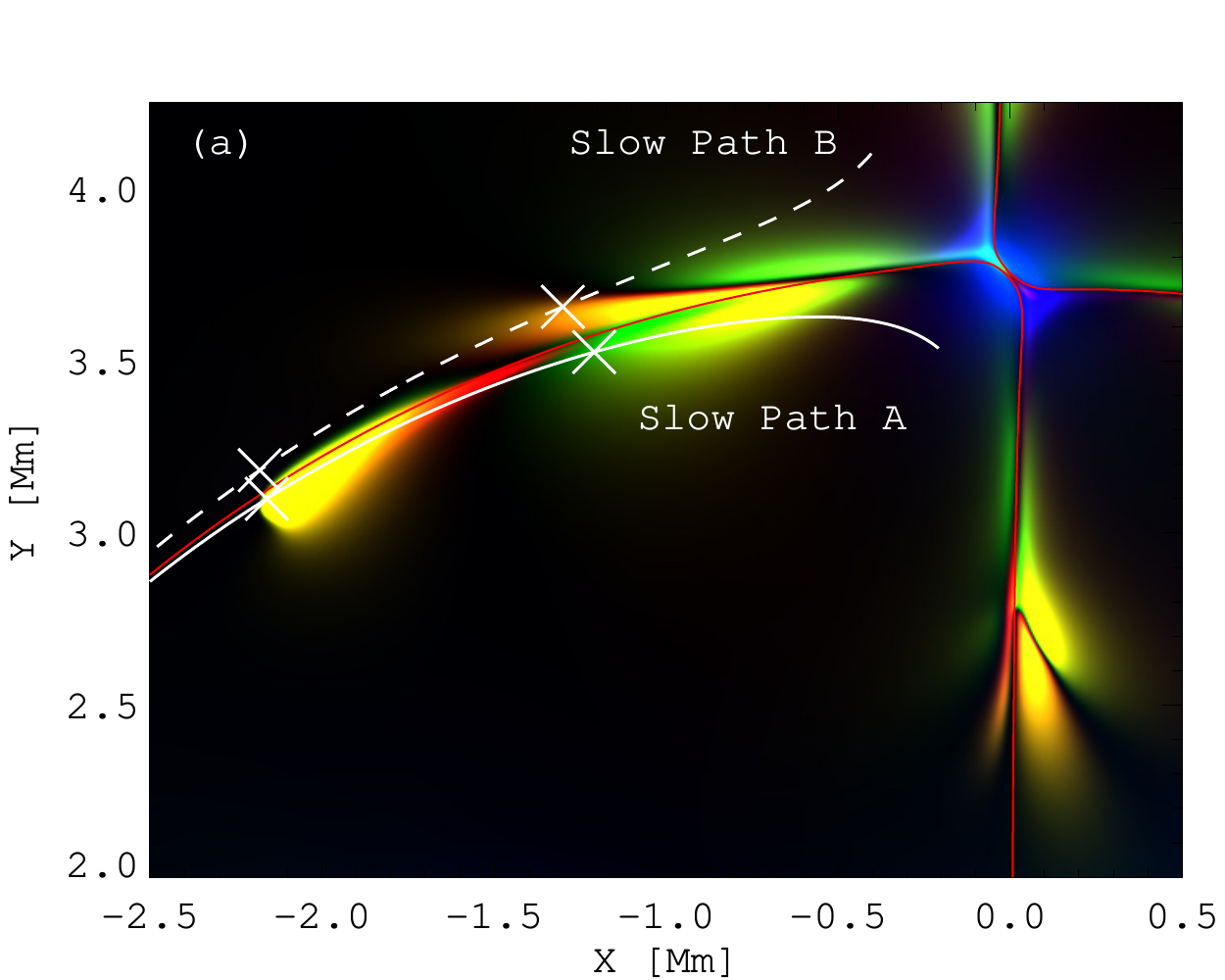}\\ \includegraphics[width=0.45\textwidth]{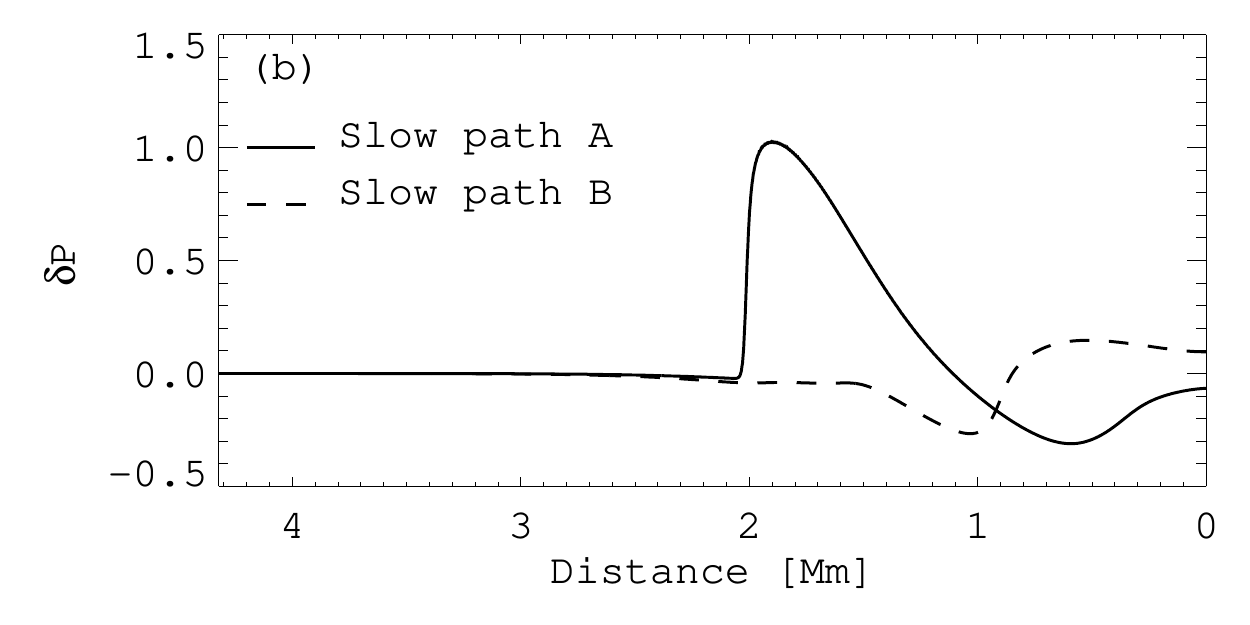}
    \caption{\label{fig:ss-asym}(a) Trajectories of two slow mode disturbances (solid, dashed white) that form on either side of the separatrices (solid red), overlaid on a wave energy density plot at time $5.7t_N$, as in \figref{fig:enfull}.  Cross marks indicate $1\unit{Mm}$ increments along each path.  (b) Fractional pressure change along each trajectory.}
  \end{figure}
\end{center}

\figref{fig:ss-asym} compares the fractional pressure change along two slow mode trajectories, $\delta P$ defined analogously to $\delta\rho$ in Equation \eqref{eq:drho}.  Panel (a) shows the wave energy densities using the same scaling as in \figref{fig:enfull}, at $t=5.7t_N$, when disturbances are present on either side of the separatrix.  Slow Path A, which is the same as that in \figref{fig:multi-csva}, traces one of the primary mode-converted disturbances, while Slow Path B follows an energy pathway that has first circled over the top of null, similar to the dashed path in \figref{fig:f2s}.  More energy follows path A than path B.  Because all of this energy eventually dissipates along each trajectory (and all corresponding trajectories near each of the other separatrices), and there is more energy along path A than path B, the regions of the simulation that experience the first and strongest mode conversions are ultimately heated more.  \figref{fig:ss-asym}a also shows the asymmetry across separatrix at $x=0\unit{Mm}$.  

The strongest shock heating is thus localized to one side of each separatrix.  The resulting pressure imbalance causes the more heated side to expand and the less heated side to contract until an equilibrium is reached between the pressure gradient and Lorentz forces.  The initial condition had a zero current (potential) magnetic field, so the new equilibrium contains currents, localized to each separatrix surface.  Thus, a tangential discontinuity forms across each separatrix.  The next two sections detail the consequences of asymmetric heating across the dome.

\section{Long term properties}\label{sec:longtermprops}
In order to better understand the interplay between current sheet formation, shock heating, and the size of the equipartition contour, we repeated a version of the simulation using the four different values for the transition region height: $y_{tr}=3.0,\ 3.15,\ 3.3$, and $3.45\unit{Mm}$.  These values generate the four different stratification profiles (and corresponding changes in the size of the equipartition region) shown in \figref{fig:multi-csva}.  For each simulation, we launched a high amplitude ($0.5v_N$) wave packet from $x_0=-1\unit{Mm}$, then allowed the system to relax.  In each case, we find that the temperature increased throughout the flux domains lying underneath the separatrix dome.  However, the enhanced temperature region becomes more strongly localized to the separatrix as the equipartition region shrinks.  

\begin{figure*}
  \includegraphics[width=\textwidth]{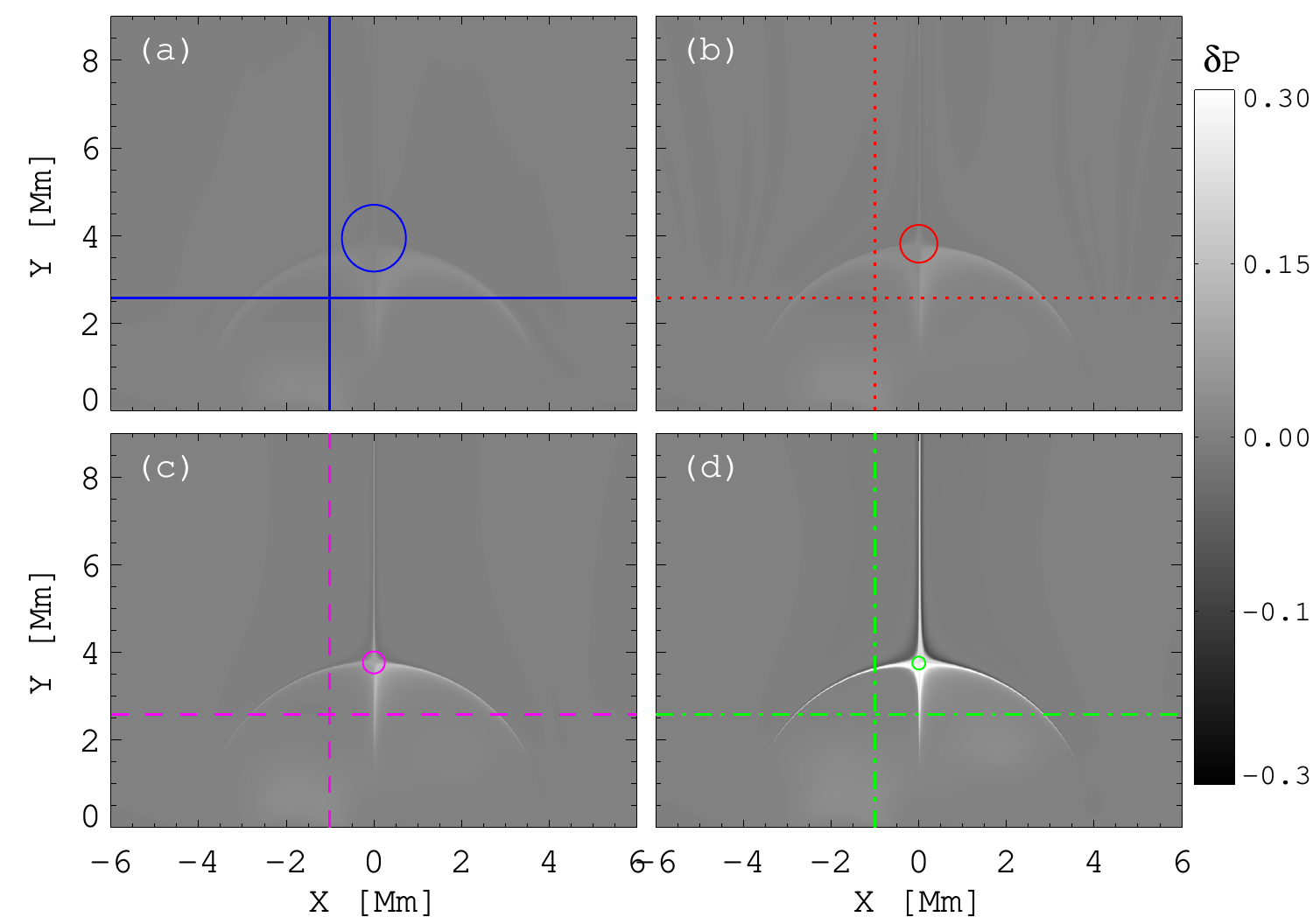}
  \caption{\label{fig:dPressure-map} Fractional pressure perturbation $\delta P$ long after single wave packet has been introduced at $x=-1.0\unit{mm}$.  (a) $y_{tr} = 3.00\unit{Mm}$, (b) $3.15\unit{Mm}$, (c) $3.3\unit{Mm}$, and (d) $3.45\unit{Mm}$.  Horizontal and vertical lines show the locations of the lineplots in \figref{fig:dPressure-line}.  The pressure perturbation peaks at $>30\%$ near the separatrices for $y_{tr}=3.45\unit{Mm}$.  The nearly--circular equipartition contours around the nulls are reproduced from \figref{fig:multi-csva}.}
\end{figure*}

\begin{figure*}
  \includegraphics[width=\textwidth]{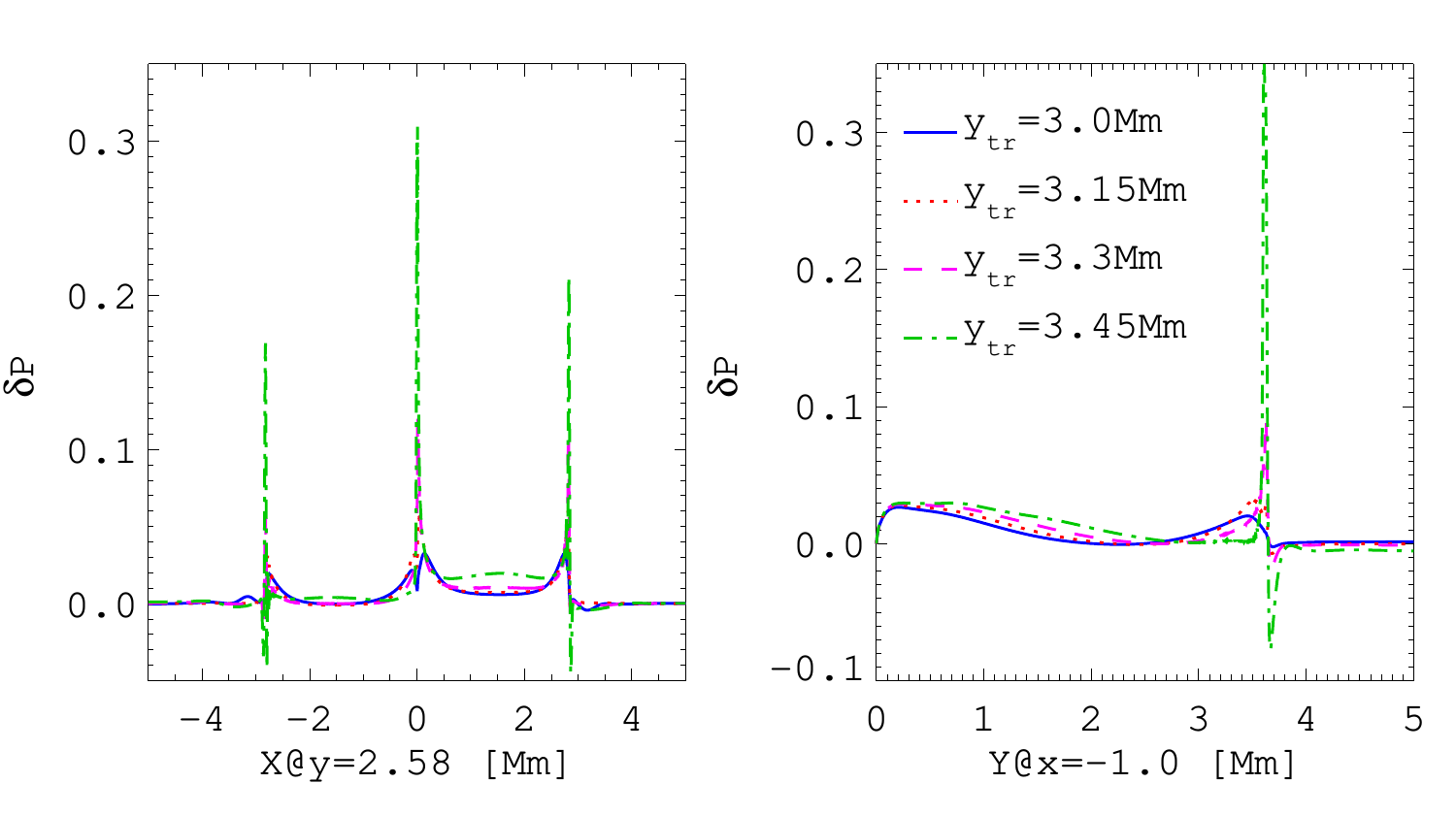}
  \caption{\label{fig:dPressure-line} Pressure perturbation along the horizontal (left) and vertical (right) slices through the domain indicated in \figref{fig:dPressure-map}.  Transition region values $y_{tr}=3.00,3.15,3.30,3.45\unit{Mm}$ correspond to solid, dot, dash, and dash--dot lines, respectively.}
\end{figure*}

\figref{fig:dPressure-map} shows a map of the pressure variance $\delta P(x,y)$.  Continuing oscillations are visible in each simulation, and the phase of these oscillations varies from map to map because of the strong dependence of the Alfv\'en speed on $y_{tr}$.  These oscillations are mostly visible outside the dome.

In contrast, the variations in plasma pressure surrounding the separatrix are basically stationary long after the pulse is introduced.  As the height of the transition region increases the pressure increase becomes more strongly localized to the separatrix, as seen in the line plots of \figref{fig:dPressure-line}.  This is due to the shrinking of the equipartition region as the transition region height is increased, with the narrower conversion region producing more tightly localized shocks near the separatrix.  Additionally, in the $y_{tr}=3.0\unit{Mm}$ case, the conversion primarily produces simple waves instead of shocks because of shallower gradients throughout the simulation, leading to comparatively less shock heating.

In all cases, energy conversion is asymmetric across each separatrix, as in the discussion of \figref{fig:ss-asym}.  The asymmetric shock dissipation across the separatrices gives the plasma a higher pressure, due mostly to enhanced internal energy, on inside of the dome.  The pressure gradient across the dome is balanced by a Lorentz force, which accounts for the enhanced current density near the separatrices.  Thus, both the magnetic and plasma properties change discontinuously across the separatrices: these are tangential discontinuities, which are allowed normal to topological boundaries \citep[see, e.g.,][\S20.2.2]{Goedbloed:2010}.

\section{Parameter study}\label{sec:param_study}

\floattable
\begin{deluxetable}{lc|cccccccccccccc}
\tablecaption{Simulation Grid Results: Fraction of injected energy to reach the null's wave conversion layer, defined in terms of the equipartition radius $r_E$ and scaleheight $H_E$, as a function of the injection location $x_0$ and the transition region height $y_{tr}$.\label{tab:simgrid}}
\tablecolumns{16}
\tablehead{
\colhead{} & \colhead{$x_0 (\unit{Mm}):$} &  \colhead{$-6.0$}&\colhead{$-5.0$} &\colhead{$-4.0$} &\colhead{$-3.0$} &\colhead{$-2.0$} &\colhead{$-1.5$} &\colhead{$-1.0$} &\colhead{$-0.5$} &\colhead{$0.0$} &\colhead{$0.5$} &\colhead{$1.0$} &\colhead{$1.5$} &\colhead{$2.0$} &\colhead{$3.0$} \\
\colhead{$y_{tr} (\unit{Mm})$} & \colhead{$r_E,\ H_e (\unit{Mm})$}   & \colhead{}   & \colhead{}   & \colhead{}   & \colhead{}   & \colhead{}    & \colhead{}   & \colhead{}    & \colhead{}   & \colhead{}    & \colhead{}   & \colhead{}    & \colhead{}   & \colhead{}
}
\startdata
3.00 & 0.77,\ 0.48   &        &   &   &   &   &  & 0.17 &    &   &    &   &    &   &  \\
3.15 & 0.45,\ 0.24   &        &   &   &   &   &    & 0.09 &    &   &    &   &    &   &  \\
3.30 & 0.28,\ 0.14   & 0.0005 &   0.0009 & 0.0006 & 0.008 & 0.02 & 0.03 & 0.04 & 0.03 & 0.0001 & 0.03 & 0.04 & 0.03 & 0.02 & 0.008 \\
3.45 & 0.19,\ 0.09   &  &   &   &   &   &    & 0.02 &    &   &    &   &    &   &  \\
\enddata
\end{deluxetable}

In the previous section we demonstrated how the development of shocks and current sheets depend on the atmospheric parameters of the system.  To study this topic further, we ran a set of simulations to form a parameter study on two variables: the height of the transition region $y_{tr}$ and the centroid location of an injected wave packet, $x_0$.  The initial condition for each of these new simulations is the near--equilibrium state reached in the four simulations with varying $y_{tr}$ described in the previous section.  To each of these simulations we inject a new wave packet with the properties described by equations \eqref{eq:vdrive} and \eqref{eq:de-drive}.  We perform 14 simulations for the $y_{tr}=3.3\unit{Mm}$ case, varying the injection location $x_0$ between $-6$ and $3\unit{Mm}$, and 3 additional simulations, injecting the wave packet at $x_0=-1\unit{Mm}$ for the cases $y_{tr}=3.0,\ 3.15,$ and $3.45\unit{Mm}$.

For each simulation, we measured the magnetic and acoustic fluxes, defined in Equation \eqref{eq:wave-conservation}, entering and exiting the null region, by repeating the analysis given in \S5.3 of \citetalias{Tarr:2017a}, e.g., using their Equations (23) for $W(t)$ and (24) for $W_\text{flux}(t)$, which are reproduced below.  The total wave energy inside a region $\mathcal{A}$ is given by the spatial integral of the wave energy density over the region at each time,
\begin{gather}
  W(t) = \int_\mathcal{A}\Bigl(E_{K}(t)+E_A(t)+E_M(t)\Bigr)d\mathcal{A},
\end{gather}
while total flow of energy into or out of the region is given by the integral of the acoustic and magnetic energy fluxes across the boundary $\partial\mathcal{A}$:
\begin{gather}
  W_\text{flux}(t) = \int_0^t \int_{\partial \mathcal{A}}\Bigl(\vect{F}_{A}(t^\prime)+\vect{F}_{B}(t^\prime)\Bigr)\cdot \hat{\vect{n}}dl dt^\prime.
\end{gather}
As in \citetalias{Tarr:2017a}, we take $\mathcal{A}$ to be the area within one equipartition scaleheight of the equipartition layer.  The equipartition layers are approximately circular with average radii $r_E$.  The equipartition scaleheight, defined in \citet{Schunker:2006}, is $H_E^{-1}\equiv\partial_\tau((c_s^2/v_A^2)\rvert_{c_s=v_A})$ where we use the boundary normal for the derivative direction $\tau$ and then average over the $c_s=v_A$ surface.  The values for $r_E$ and $H_E$ for each value of $y_{tr}$ are given in Table \ref{tab:simgrid}.

\begin{figure}
  \includegraphics[width=0.45\textwidth]{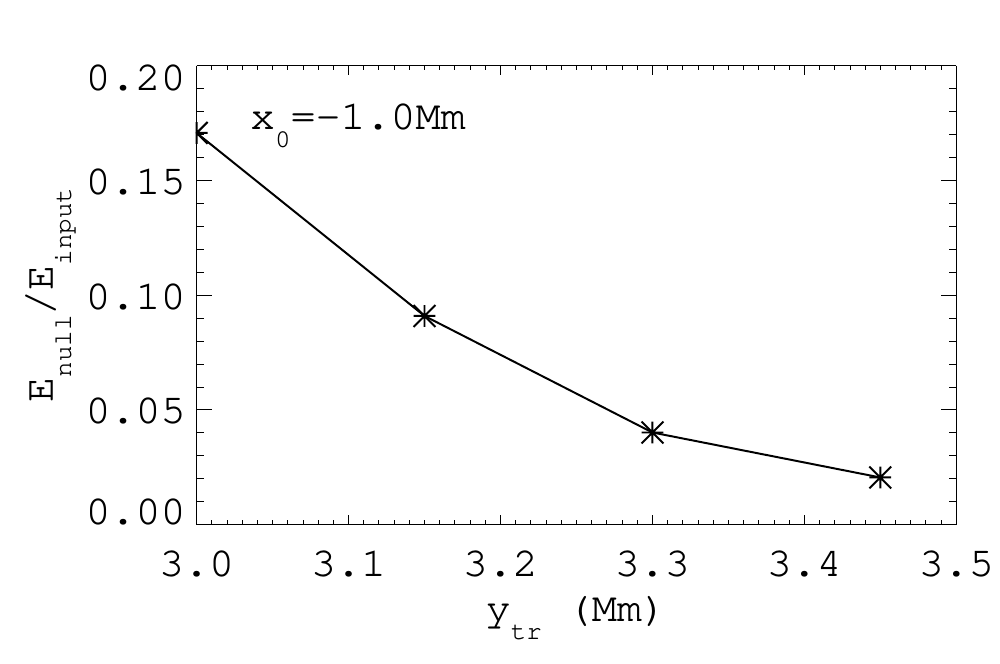}
  \caption{\label{fig:enull-vary-ytr} Fraction of energy of an injected wave packet to reach the null point region as the transition region height is varied: $y_{tr}=3.00,3.15,3.30,\ 3.45\unit{Mm}$.  }
\end{figure}

\begin{figure}
  \includegraphics[width=0.45\textwidth]{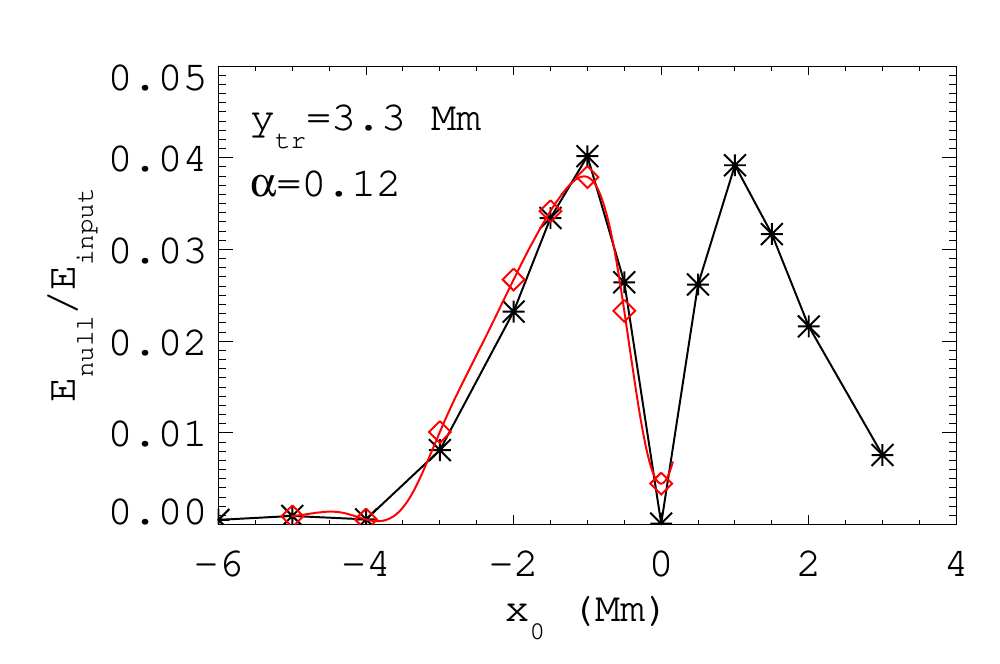}
  \caption{\label{fig:enull-vary-x} Fraction of energy of an injected wave packet to reach the null for $y_{tr}=3.3\unit{Mm}$, as the injection location $x_0$ is varied (asterisks, black line).  The WKB estimate from \citetalias{Tarr:2017a} Fig.~11, scaled by a factor $\alpha\approx 0.12$, is shown in red.  This scaling minimizes the square difference between the WKB estimate sampled at the simulation points (diamonds) and the results of the \change{$y_{tr}=3.3\unit{Mm}$} simulations.}
\end{figure}

The total energy of the injected wave packet, $E_\text{input}$, is known from the simulation setup.  The total energy density flux through the equipartition region around the null is then used to measure the fraction of the initial injected wave energy that makes it to the null.  The ratio of the net acoustic flux to net Poynting flux which passes through the equipartition contour surrounding the null measures the efficiency of mode conversion in that region, thus giving the fraction of the initial injected energy that ends up as the slow mode shocks along the separatrices.  The results for this analysis for the 17 simulations are summarized in Table \ref{tab:simgrid} and plotted in Figures \ref{fig:enull-vary-ytr} and \ref{fig:enull-vary-x}.  We found the mode--conversion efficiency varied only by $\approx 20\%$ across all simulations, much less than the other factors, and do not consider it further.

\figref{fig:enull-vary-ytr} shows that the fraction of injected wave energy to reach the null decreases as the height of the transition region increases.  This quantifies the discussion in \S\ref{sec:longtermprops}.  The result for $y_{tr}=3.0\unit{Mm}$, with \change{$E_{\text{null}}/E_\text{{input}}\approxeq 0.17$}, repeats the result of \citetalias{Tarr:2017a}\footnote{The fraction given here of $0.17$ differs slightly from the $0.155$ reported in the previous work, due to the slightly different background conditions.}.  A ratio of $0.04$ was found for the primary simulation discussed in the present work, where $y_{tr}=3.3\unit{Mm}$.

\figref{fig:enull-vary-x} shows the effect of varying the wave packet injection location $x_0$ for the $y_{tr}=3.3\unit{Mm}$ simulations.  These results are shown as asterisks connected by solid black lines.  In \citetalias{Tarr:2017a} \S6 we calculated the effect of varying the wave packet injection location using a WKB method, (see Figure 11, dot-dashed line of that paper).  The WKB estimate agrees with the simulation results when scaled by a factor $\alpha\approx0.12$.  That factor, which minimizes the square--difference between the simulation results and the WKB estimate, is expected: it accounts for both the result of \citetalias{Tarr:2017a}, that the WKB method overestimates the amount of energy to reach the null by a factor of 1.8, and for the results of \figref{fig:enull-vary-ytr}, that a factor 4.25 \emph{less} energy makes it to the null for $y_{tr}=3.3\unit{Mm}$ compared to the $y_{tr}=3.0\unit{Mm}$ case.

The strength of the outgoing shocks depend on the size of the equipartition region near the null and the amount of the wave's initial energy that makes it to that region (in addition to the initial injection amplitude).  In our simulations, as the height of the transition region increases and the equipartition region shrinks, the amount of energy that makes it to the null also decreases.  At the same time, the null collapse and conversion process itself becomes more violent and localized, leading to stronger shock formation.  This results in the localization of the pressure disturbance seen in Figures \ref{fig:dPressure-map} and \ref{fig:dPressure-line}.

\section{Discussion}\label{sec:discussion}
The results of the previous two sections have important implications for heating regions of the Sun that share this type of dome topology, especially coronal bright points \citep[CBP:][]{Golub:1974,Alipour:2015}.  CBPs are typically (though not always) associated with a magnetic null and dome topology \citep{Galsgaard:2017}.  From our study, it is clear that fast mode waves introduced at different locations throughout a dome (and even slightly outside a dome) will refract towards the null, undergo mode conversion, and form slow shocks localized to the separatrices.  If we assume that the incident wavefield is stochastic---i.e., not a single wave packet localized to one lobe of the magnetic dome, but a continuous, uniform spectrum across the entire lower boundary, then the average of the curve in \figref{fig:enull-vary-x} is the expected energy to be pumped from the wave field into shocks and Ohmic dissipation, and ultimately heating the plasma in the dome.  That value is 0.016, meaning that of order $1\%$ of the high frequency\footnote{\change{\citep[Say, a few times the Br\"unt--V\"ais\"al\"a or acoustic cutoff frequency, both of which are $\approx 5\unit{mHz}$ in the photosphere or low chromosphere][]{Mihalas:1981, Vigeesh:2017, Cally:2006}.}} convective wave flux through the photosphere could heat the solar plasma in these regions.

It is interesting to consider whether the processes described in this paper may help explain the association of CBPs with magnetic dome topologies.  If the magnetic concentrations associated with them do not substantially suppress convection, then the convective wave field will continually pump energy into the plasma near the separatrices.  For a survey of 285 CBPs, \citet{Longcope:2001a} determined their radiated power using SOHO EIT and MDI data \citep{Delaboudiniere:1995, Scherrer:1995}.  From the distribution of radiated power versus CBP size they found a typical heat flux over the CBP area of $1000\unit{W}\unit{m}^{-2}$.  A typical radius for these structures is $5\unit{Mm}$, similar to the size of our simulated dome.  \citet{BelloGonzalez:2010} found that the total upward acoustic flux through the photosphere to be $\approx 7000\unit{W}\unit{m}^{-2}$; this is roughly double previously reported values, owing to better spatial and temporal coverage.  Those previous estimates \citep{BelloGonzalez:2009} further found that $\approx 1/3$ of the total power is contained in the high frequency component $(>10\unit{mHz})$, or $\approx 2000\unit{W}\unit{m}^{-2}$ using the newer total flux estimates.  These measurements then indicate that if high frequency waves are responsible for CBP emission, $50\%$ of the high frequency energy incident upon a CBP dome must be converted to thermal energy within the dome to account for the observed CBP heat flux of $1000\unit{W}\unit{m}^{-2}$.  Instead, we found of order $1\%$ efficiency, so the process we have outlined above cannot be the main source of power necessary to replenish radiative losses by CBPs that have a dome--type magnetic topology.  Under the most favorable conditions from our parameter study, the focusing of wave energy towards the null to generate shocks and current sheets might account for $5-10\%$ of the the radiative loss budget of CBPs.  Apparently, the bulk of the energy must come from other processes, such as reconnection due emergence or continual shearing, as discussed in \citet{Galsgaard:2017}.

The type of wave dissipation we have studied is also unlikely to play a dominant role in larger structures.  \citet{Tarr:2014} determined the total radiated power from an emerging ephemeral region, again with a dome topology similar to the one we have simulated here.  That region contained a coronal null point $\approx 15\unit{Mm}$ above the photosphere whose fan surface intersected the photosphere in a ring roughly $30-50\unit{Mm}$ across, covering an area $\approx10^{15}\unit{m}^2$.  \change{They found} a dome--shape structure in many EUV and X--Ray wavelengths that clearly coincides with the null's separatrix surface (c.f. their Figures 2,3,5, and 7).  The dimensions for this structure are rather larger than our simulated dome rotated about its central axis.  The total radiated power from this region varied between $10^{18}-10^{19}\unit{W}$ during emergence, giving a radiated flux of some $2000-4000\unit{W}\unit{m}^{-2}$ from the dome, essentially the same as for CBPs.  Once again, clearly another process, such as the reconnection between emerging and external fields proposed by \citet{Tarr:2014}, must play a more dominate role in that case.

On the other hand, the formation of the upward propagating slow shocks in our simulation may help explain what drives type--II spicules \citep{DePontieu:2007}.  \citet{Martinez-sykora:2017a}, particularly in their supplementary material, presented a simulation of the chromosphere that includes sets of transverse waves that mode convert into slow mode waves and associated flows, which they identify as type--II spicules (see their Figure S6 and its caption).  At locations where they identify the conversion process, both the field line structure and the existence of closed--contours of the $\beta=1$ surface point to the presence of magnetic X--lines or null points: see, for example, their Figure S3 at location $\approx (91.5,2\unit{Mm})$, or Figure S6 at location $(22,2\unit{Mm})$.  Furthermore, out--of--plane currents localize at the approximate locations of the separatrices attached to the conversion points, similar to the concentration of currents we find in our simulations (see, e.g., their Figure S3, especially panel D).  This suggests that a similar mechanism leads to both their results and ours, even though their simulations include the additional physics of ambipolar diffusion.

The slow mode shocks we find at first seem similar to those reported in \citet{McLaughlin:2009}.  They, too found slow mode shocks traveling outward from \change{a null} after a fast mode wave passed through.  However, these shocks arise from a different process.  Their initial condition is zero $\beta$ so only fast modes are initially supported by the plasma.  An incoming initially symmetric pulse eventually creates a ring of fast shocks.  The shocks collide, forming a cusp, and the heating is much greater in the cusps.  The pressure gradient set up in the cusps then gives rise to a set of outward propagating slow shocks.  The inward propagating fast shocks then collapse the null into a current sheet and drive reconnection outflows (jets) into the cusp, further heating the plasma and driving the slow shocks.  

Our slow shocks form instead by mode conversion of the fast wave, a process not immediately available when the initial condition is zero $\beta$.  This difference is illustrated by considering the low amplitude limit.  When \citet{McLaughlin:2009} reduced their initial amplitude to keep the simulation in the linear regime \change{(see their Appendix A)}, the slow shocks disappear from the solution while the fast mode wave simply propagates symmetrically inward with increasing energy density.  In contrast, in our simulations if we reduce the amplitude of the perturbation or change the system parameters $(y_{tr})$ to make the gradient of the phase speed more gradual, the strong slow mode shocks smoothly transition into weak shocks and then slow mode waves, still localized to the separatrices, which are what we reported in \citetalias{Tarr:2017a}.

Does this analysis carry over into 3D?  \citet{Thurgood:2017} shows that yes, it can for an idealized 3D null.  They perform a 3D upgrade to \citet{McLaughlin:2009}, where the initial pulse is \change{a} donut--shaped fast wave, and they find very similar propagating density shocks to those we discussed in \S\ref{sec:shocks} (see their Figure 8).  This demonstrates that the conversion of incoming waves to compressive modes remains a substantial effect in 3D.  On the other hand, they also find evidence that the incoming fast wave generates outward propagating Alfv\'en waves, which we deliberately avoided in our 2D setup.  Clearly in 3D, in order to understand how the energy of propagating waves is ultimately distributed, one must treat the conversion between all modes simultaneously.  This will be the basis of a future study.

Another further step into the 3D realm has already been taken by \citet{McLaughlin:2019}, who performed a 3D WKB study of fast wave refraction in a 3D dome topology.  They could not address mode conversion due to their cold plasma approximation, by they did determine that the ability of a null to directly capture wave energy decreases as the null height increases above the photosphere.  At the same time, it is apparent that any inhomogeneity in the magnetic field affects the propagation of MHD waves.  Relaxing the $\beta=0$ assumption and adding a stratified atmosphere are important next steps in determining how waves propagate in solar--like environments.

Santamaria and coauthors have discussed a similar set of simulations in a series of papers \citep{Santamaria:2015, Santamaria:2016, Santamaria:2017,Santamaria:2018}.  \citet{Santamaria:2017} report on a set of jet--like slow shocks leaving a null, analogous to those we analyzed in detail.  They focus mostly on properties of the wave field induced by the wave--null interaction, finding that the null may act like a resonant cavity, producing higher frequencies compared to those originally introduced in their wave packet.  As they further report in \citet{Santamaria:2018}, the high frequency disturbances appear to be localized to the separatrices, so that the high frequencies they find are related to the properties of mode-converted slow mode shocks: higher amplitude, faster traveling shocks produce higher frequency dynamics.

We do not appear to find direct evidence of a resonant cavity at the null in our simulations.  The apparent oscillatory reconnection we find seems to be related to the properties of the incoming wave (reconnection once in each direction corresponding the leading and trailing edges of the pulse, then repeated for each reflection off the lower boundary).  Oscillatory reconnection has been considered in other studies \citep{Craig:1991, Murray:2009, McLaughlin:2012a, McLaughlin:2012b, Threlfall:2012, Thurgood:2019}.  Notably, in these simulations the region about the null is either continuously driven, say by ongoing flux emergence \citep{Murray:2009, McLaughlin:2012b}, or the initial perturbation has a strong symmetry about the null while the resulting dynamics include reflections off exterior boundaries \citep{Craig:1991, McLaughlin:2012a, Threlfall:2012, Thurgood:2019}.  Oscillatory reconnection should be a process inherent to the dynamics of the null, but appears difficult to isolate in all of the above studies.  Still, the nonlinear dynamics reported in \citet{Thurgood:2019} belie a simple explaination due to reflections off a boundary.  That we fail to see oscillatory reconnection in our simulation could be because: (i) our resistivity \change{of $0.0333$} is too high and therefore creates too much damping (the inherent length scale of a stratified atmosphere leads to rather high numerical resistivity when attempting to resolve both a transition-region null and a photosphere, see \citet{Tarr:2017a} \S2.1); (ii) the efficiency of mode conversion allows the slow modes to carry away the energy that would otherwise act as the potential energy needed to cause rebound; or (iii) the wave packets that reach the null do so rather asymmetrically, while a \change{quadrupolar (cartesian) mode about the null may be necessary to excite oscillatory reconnection \citep{Craig:1991, Longcope:2012}.}

Our simulations neglected radiation, but the general effect has long been understood \citep[][\S6.5.3]{Bray:1974}.  Shocks cause nonadiabatic changes: they increase the plasma entropy, while radiation removes entropy from the local plasma (indeed, plots of the change in plasma entropy $S=P\rho^{-\gamma}$ for our simulations show similar structure to \figref{fig:dPressure-map}).  A full shock cycle including radiation will return the plasma to its initial state.  If we included radiative losses, the shocks in our simulation would lose energy faster by radiative damping of the shock front, while the tangential discontinuities in the shock wake will reduce with time as the heated plasma radiatively cools and the balance between plasma pressure gradients and Lorentz forces shifts back towards a more magnetic--potential state.  Thus, for solar observations, we would expect the mode--conversion/shock process to involve a more localized intensity enhancement that quickly fades.  The upward and downward propagating shocks would generate some Doppler shifts of spectral lines, but only as a component of the background emission.  This behavior is similar to the observations reported in \citet{Reardon:2013}, who found localized, spatially spreading intensity enhancements, $10-20\%$ above the background, and perhaps $1-1.5\unit{Mm}$ at maximum extent.  However, further simulations that include radiation are required for more detailed comparison.

\section{Conclusion}\label{sec:conclusion}
We have studied the dynamics that arise when compressive waves are introduced in stratified, solar--like atmospheres that include magnetic nulls.  A surprising result of our previous work \citepalias{Tarr:2017a} was that purely compressive waves introduced in the photosphere lead to currents localized to the magnetic skeleton of the overlying field.  Here, we report the mechanism by which that localization occurs.  We found that the initial current sheet formation is dominated by magnetic compression localized to one side of the initial null point.  Magnetic reconnection across this current sheet locally heats the plasma near the null through Ohmic dissipation.  There are two distinct reconnection scenarios at work.  The first is caused by fast mode refraction and subsequent conversion around null region, and is responsible for the majority of the Ohmic heating.  The second appears to be due to diffusion as the fast wave reflects off the lower boundary, and thus does not substantially contribute to the heating at the null.  The resulting plasma pressure gradients are balanced by the magnetic pressure term in the Lorentz force, leading to a persistent current density localized to the null.

The currents along the separatrices, on the other hand, are caused by mode--converted slow mode waves that shock as they propagate away from the null.  The strongest shocks form underneath and adjacent to the separatrix dome, so the plasma is heated more in those locations compared to the opposite sides of the separatrices.  As the higher pressure regions expand they compress the magnetic field on the opposite side of the separatrix, generating the long--lived currents localized to each separatrix.

The fraction of the injected wave packet's energy which eventually heats the plasma depends on the region of the dome where the packet is introduced and the properties of the stratified atmosphere.  We found that our previous \citepalias{Tarr:2017a} WKB estimate of this dependence matched our full MHD simulation results quite well.  A greater decrease in density with height, caused by a higher transition region, allows less wave energy to reach the null, while a more compact equipartition region surrounding the null generates more focused mode conversion and stronger slow mode shocks along the separatrices, better localizing energy dissipation to the separatrix surface.

Finally, we applied our results to the question whether the trapping and dissipation of high frequency waves in magnetic dome topologies association with coronal bright points is sufficient to balance observed radiative losses.  We found that the processes we described may only account for $\lesssim10\%$ of the required enegies under the most favorable circumstances.

\acknowledgements
This work is supported by the the Chief of Naval Research and the NASA HSR and LWS programs.  The simulations were performed under a grant of computer time from the DoD HPC program.  This research has made use of NASA's Astrophysics Data System.  A portion of this work was carried out while L.A.T. was a visiting scientist at the University of Hawaii Institute for Astronomy, Maui, HI, and as an NRC Research Associate hosted by the Naval Research Laboratory in Washington, DC.  We thank an anonymous referee for many helpful comments.

\software{LARE2D v.2.11 \citep{Arber:2001}}

\appendix
\section{Grid stretching}\label{sec:stretch}
The stretched grids are defined in cell coordinates and then scaled to the total simulation length.  In the $x$ direction, $L_\text{total}=700L_N$, and the width of a cell $-N_x/2\leq ix\leq N_x/2$ is given by
\begin{equation}
  dx(ix) = dx_0 + dx_0\frac{f_x}{2}\Bigl[2-\tanh\frac{ix-c_1}{w}+\tanh\frac{ix-c_2}{w}\Bigr]
\end{equation}
where we use $dx_0 = L_N/32,\ c_1 = -N_x/2+512,\ c_2 = N_x/2-512$, and $w=160$.  The above function smoothly varies the cell width between $dx_0$, or $\Delta_x=L_n/32$, at the center of the domain (near the magnetic dome) to $\Delta_x=dx_0(1+2f_x)$, or $\Delta_x=0.526L_N$, at the simulation edges.  The geometric factor is $f_x/2=\frac{L_\text{total}/dx_0-N_x}{N_x+A_x}$, where $A_x=2w\ln\Bigl(\cosh\frac{N_x/2-c_1}{w}\sech\frac{-c_2}{w}\Bigr)$.  The factor $f_x$ is related to the amplitude of the stretching, while $A_x$ is related to the fraction of stretched versus unstretched cells.

The $y$ direction is defined similarly, where now $L_\text{total}=800L_N$, $0<iy<N_y$, and 
\begin{equation}
  dy(iy) = dy_0 + dy_0f_y\Bigl[1+\tanh\frac{iy-c}{w}\Bigr],
\end{equation}
in which $dy_0 = L_N/32,\ c = 1920$, and $w=160$.  The $y$ direction has only one stretching center, so the geometric factors are slightly different: $f_y = \frac{L_\text{total}/dy_0-N_y}{N_y+A_y}$ and $A_y=w\ln\Bigl(\cosh\frac{N_y-c}{w}\sech\frac{-c}{w}\Bigr)$.  The $y$ grid varies smoothly between $\Delta_y=dy_0=L_N/32$ and $\Delta_y=dy_0(1+2f) = 0.34L_N$.

\end{document}